\documentclass[aps, prx,twocolumn,longbibliography,superscriptaddress,amsmath,amssymb,floatfix]{revtex4}
\usepackage[latin9]{inputenc}
\usepackage{amssymb}
\usepackage{graphicx}
\usepackage{amsmath}
\usepackage{color}
\usepackage{mathrsfs}
\usepackage{float}
\usepackage{indentfirst}
\usepackage{mathrsfs}
\usepackage{float}
\usepackage{indentfirst}
\usepackage{textcomp}
\usepackage{comment}
\usepackage{mathtools}
\usepackage{natbib,hyperref}
\usepackage{soul}

\begin{document}

\title{From Tree Tensor Network to Multiscale Entanglement Renormalization Ansatz }

\author{Xiangjian Qian}
\affiliation{Key Laboratory of Artificial Structures and Quantum Control (Ministry of Education),  School of Physics and Astronomy, Shanghai Jiao Tong University, Shanghai 200240, China}

\author{Mingpu Qin} \thanks{qinmingpu@sjtu.edu.cn}
\affiliation{Key Laboratory of Artificial Structures and Quantum Control (Ministry of Education),  School of Physics and Astronomy, Shanghai Jiao Tong University, Shanghai 200240, China}

\begin{abstract}
    Tensor Network States (TNS) offer an efficient representation for the ground state of quantum many body systems and play an important role in the simulations of them.
    Numerous TNS are proposed in the past few decades. However, due to the high cost of TNS for two-dimensional systems, a balance between the encoded entanglement and computational complexity of TNS is
    yet to be reached. In this work we introduce a new Tree Tensor Network (TTN) based TNS dubbed as Fully-Augmented Tree Tensor Network (FATTN) by releasing the constraint in Augmented Tree Tensor Network (ATTN). 
    When disentanglers are augmented in the physical layer of TTN, FATTN can provide more entanglement than TTN and ATTN. At the same time, FATTN maintains the scaling of computational cost with bond dimension in TTN and ATTN.
    Benchmark results on the ground state energy for the transverse Ising model are provided to demonstrate the improvement of 
    accuracy of FATTN over TTN and ATTN.
     Moreover, FATTN is quite flexible which can be constructed as an interpolation between Tree Tensor Network and Multiscale
    Entanglement Renormalization Ansatz (MERA) to reach a balance between the encoded entanglement and the computational cost.    
     %To date, we have a lot of methods in Tensor Network States, including Density Matrix Renormalization Group (DMRG), Projected Entanglement Pair States (PEPS), Multiscale Entanglement Renormalization Ansatz (MERA), Tree Tensor Network States (TTN). However,  most of the methods have their own drawbacks. In order to efficiently simulate quantum many body systems, we propose a new Tensor Network structure, namely Fully-Augmented Tree Tensor Network(FATTN), based on Augmented Tree Tensor Network(ATTN) to simulate the high-dimensional quantum many body systems. By adding disentanglers not sharing the same physical index, FATTN can achieve a more entangled Tensor Network state than ATTN and simultaneously maintain a low computation complexity of $O(D^4d^4)$.  As a benchmark for our ansatz, we calculate the ground state energy for the transverse Ising model on a square lattice with periodic boundary conditions, providing the most accurate results compared with TTN and ATTN. 

\end{abstract}

\maketitle

\section{Introduction}
Understanding exotic phases and exotic phase transitions in strongly correlated quantum many body systems
is one of the most challenging topics in condensed matter physics \cite{2012LNP...843.....C,Xiao:803748,marino_2017}. 
Due to the lack of analytic solutions, most studies of these systems depend on many-body numerical approaches \cite{PhysRevX.5.041041}.
For quantum many body systems, the exponential increase of the dimension of the Hilbert space prevents us from
studying system with large sizes. Quantum Monte Carlo is an efficient method for quantum many body systems, but it usually suffers from the
infamous minus sign problem for Fermionic systems \cite{PhysRevB.41.9301,PhysRevLett.94.170201}. 
%Tensor Network States \cite{10.21468/SciPostPhysLectNotes.8,ORUS2014117,Bridgeman_2017} provides
%an efficient way to represent many body states which can get rid of the minus sign problem.
%In quantum many body systems, the number of coefficients required to specify the quantum state grows exponentially
% with the system size. So, it's stiff baffling to deal with quantum many body problems. Fortunately, we actually don't
% need to consider the whole Hilbert space as most of the Hamiltonian we encountered in nature tend to be local. More
 % specifically,
For a local Hamiltonian, the ground state and low energy excited states
satisfy the well-known entanglement-entropic area law \cite{PhysRevLett.94.060503,PhysRevLett.90.227902,PhysRevLett.71.666,RevModPhys.82.277},
asserting that if we divide the studied system into two parts, the entanglement entropy is proportional to the
measure of the partition's boundary rather
than its volume. Hence, if we are only interested in the low-lying states, we only need to consider states which satisfy
the entanglement-entropic area law in the Hilbert space. Tensor Network States (TNS) \cite{10.21468/SciPostPhysLectNotes.8,ORUS2014117,Bridgeman_2017} can efficiently encode the entanglement-entropic area law
by design, which can faithfully represent the ground state of quantum many body systems. In recent
 years, significant advances in TNS have been made and they are becoming one of the most popular approaches in studying
strongly correlated quantum many body systems \cite{Zheng1155,PhysRevLett.118.137202}. During the last three decades, after realizing the underlying wave-functions in Density Matrix Renormalization
Group (DMRG) \cite{PhysRevB.48.10345,PhysRevLett.69.2863,cmp/1104249404,doi:10.1146/annurev-conmatphys-020911-125018,RevModPhys.77.259} are actually Matrix Product States (MPS) \cite{PhysRevLett.75.3537,SCHOLLWOCK201196},
several types of TNS for higher dimensional systems are proposed, including Tree Tensor Network (TTN) \cite{PhysRevB.80.235127,PhysRevA.74.022320,PhysRevA.81.062335,PhysRevB.90.125154},
Projected Entanglement Pair States (PEPS) \cite{RevModPhys.93.045003,2021arXiv211012726V,Verstraete:2004cf,PhysRevA.70.060302,PhysRevLett.96.220601} and
their generalization \cite{PhysRevX.4.011025},  and
 Multiscale Entanglement Renormalization Ansatz (MERA) \cite{PhysRevLett.101.110501,PhysRevLett.102.180406,PhysRevLett.99.220405,PhysRevB.79.144108}.

For (quasi) one dimensional systems, DMRG or MPS based approaches
%which can be interpreted as Matrix Product States(MPS)
 \cite{SCHOLLWOCK201196,cmp/1104249404} are now the workhorse as it can capture the one dimensional entanglement-entropic area
 law (with a logarithmic correction for critical systems) with a relatively low computational complexity.
 % We have got some considerable  results using DMRG in one-dimensional systems. 
 However, to study two-dimensional systems with DMRG, the bond-dimension needs to be increased exponentially with the width of the system to
 be able to capture the entanglement-entropic area law, which makes the study of wide system difficult with DMRG \cite{PhysRevB.49.9214}.
 %A So, it won't be an efficient way for us to simulate two-dimensional systems. 
 %We need to find some other ways to deal with two-dimensional systems. Until now, the most popular tensor network structures for two-dimensional systems are PEPS, MERA and TTN. 
 PEPS is a straightforward generalization of MPS to two dimension. It can capture the area-law of entanglement entropy for two-dimensional
 systems by design. However, it suffers from a high computational complexity which is usually higher than $O(D^{10})$
  \cite{PhysRevLett.101.090603,PhysRevA.86.022317}, with $D$ the bond-dimension of PEPS. Moreover, the overlap of PEPSs can't be calculated exactly \cite{PhysRevLett.98.140506}, even though there exist many approaches to calculate it approximately \cite{doi:10.1143/JPSJ.65.891,PhysRevB.86.045139,ORUS2014117,2021arXiv211012726V}. 
  %{\color{red}However, PEPS can directly work on thermodynamic limit.}
  On the contrary, MERA which is constructed by unitary tensors \cite{foot2}, can be contracted exactly but suffers from
  a much higher computational complexity for two dimensional systems $O(D^{16})$ \cite{PhysRevLett.102.180406}. 
  TTN has a similar structure to MERA. It has a lower computational complexity for two dimensional systems (O$(D^4)$), but encounters the same problems as the DMRG in 2D
  because TTN doesn't encode the two-dimensional entanglement-entropic area law by design. 
  So it is desirable to develop new tensor networks with affordable computational complexity which can also encode high
  entanglement.

Recently, a new tensor network structure based on TTN: Augmented Tree Tensor Network (ATTN) was proposed
in \cite{PhysRevLett.126.170603}. By placing disentanglers at the physical  layer of a TTN, it was claimed that ATTN can
capture the entanglement-entropic area-law and at the same time keep a computational complexity of $O(D^4d^2)$, with $d$ the dimension of the physical degree of freedom. 
%However, the entanglement-entropic area scale in ATTN as $L\log(2)+log(D)$ which
%means the increase of entanglement entropy with bond dimension D is still logarithmic. 

We propose a new structure which can provide larger entanglement but maintain
the low computational complexity in this work. In ATTN, no two disentanglers share
the same Hamiltonian term. But we find this is not a necessary constraint to maintain a low computational cost. We can augment TTN with disentanglers
in a way that no two disentanglers share the same physical index.   
With this strategy, we generate a Fully-Augmented Tree Tensor Network (FATTN) which can capture the entanglement-entropic area 
law of $L\log(d^2)$ (for most cuts on a finite lattice) with a large bond dimension $D$ \cite{foot3} for two-dimensional quantum many body systems on
a finite lattice and keep a relatively low computational complexity of
 $O(D^4d^4)$. 
 {So the entanglement captured in FATTN is at least as large as that in a PEPS with bond-dimension $D = d^2$. 
 	But as we will discuss late, while iPEPS can directly deal with system in the thermodynamic limit, FATTN is suitable for finite lattice with periodic boundary conditions.
 	The optimization and contraction of FATTN is easier than PEPS.
 	} 
 In our benchmark calculation of a $8\times8$ transverse Ising model near the critical
  point, FATTN gives a relative error of ground state energy of the order of $10^{-5}$ even for $D=50$ which largely improve the
  accuracy over TTN
   \cite{PhysRevB.80.235127} and ATTN \cite{PhysRevLett.126.170603}. 

The rest of the paper is organized as follows: In Sec.~\ref{sec_TTN} and Sec.~\ref{sec_ATTN}, we briefly review the TTN and ATTN. 
We then introduce FATTN and discuss the optimization of it in Sec.~\ref{sec_FATTN}. 
Benchmark results are provided in Sec.~\ref{sec_benck}. In Sec.~\ref{sec_multi}, we discuss different strategies to place disentanglers
in FATTN. We summarize this work in Sec.~\ref{sec_con}.

\section{Tree Tensor Network}
\label{sec_TTN}

\begin{figure}[t]
	\includegraphics[width=60mm]{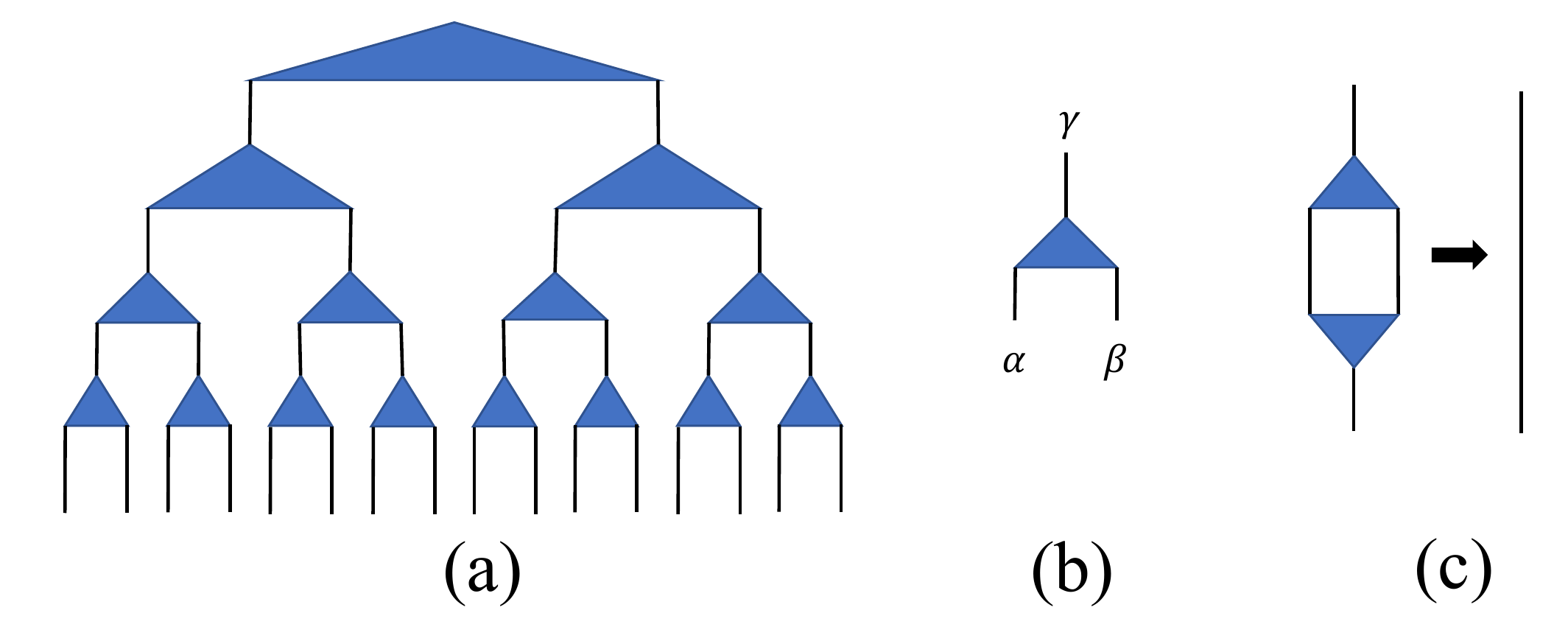}
	\caption{The structure of TTN. In (a), we show an illustration of one dimensional TTN. The building block
		of TTN: the isometry is shown in (b). In (c), we show that the contraction of an isometry
		and its conjugate gives an identity tensor from its definition in Eq.~\eqref{cons_iso}.}
	\label{TTN}
\end{figure}

%{\em Model and Methods --}
TTN is constructed by placing tensors on a tree structure \cite{10.21468/SciPostPhysLectNotes.8}. Because it is loop-free, TTN
can be efficiently contracted with a computational complexity of $O(D^4)$. However, at the worst case, it can only provide
log$(D)$ entanglement entropy similar as DMRG which means TTN can't efficiently capture the entanglement-entropic area law
of high dimensional systems.
To be self-contained, we will give a brief description on the optimization and contraction of TTN following \cite{PhysRevB.80.235127},
which are the basic steps in approximating the ground state of quantum many systems with TTN.
%In order to give the reader a pragmatic example and provide convenience for our later discussions, we will analyze the computational complexity in the process of contracting and optimizing TTN defined in the Ref. \cite{PhysRevB.80.235127}. 

As shown in Fig.~\ref{TTN}, in TTN, a quantum state $|\psi\rangle$ is represented as a contraction of a set of
isometries $w_{i}$ \cite{foot1} which are rank-3 tensors with bond-dimension $D$. (For simplicity, we assume that all the bond-dimensions in TTN are the same). The definition of isometry is
\begin{equation}
	\sum_{\beta_1\beta_2}(w_{\beta_1\beta_2}^\alpha)(w_{\beta_1\beta_2}^\alpha)^*=\delta_{\alpha,\alpha^*}
	\label{cons_iso}
\end{equation}
(The root tensor is actually a rank-2 tensor, but we can treat it as a rank-3 tensor with the dimension of the
third index $1$.)
%For some reasons, we ignore the differences between the rank-2 tensors at the top of TTN and rank-3 tensors). 
The relationship in Eq.~\eqref{cons_iso} is very useful as they can significantly reduce the computational cost when we perform tensor-tensor contraction in the optimization
of TTN and the calculation of physical observables. 

In this work, we employ a standard optimization algorithm described in Ref. \cite{PhysRevB.79.144108,PhysRevB.80.235127}
(For other optimization strategies, we refer the reader to Ref. \cite{PhysRevB.79.144108,PhysRevResearch.3.023148}).
To approximate the ground state of a given Hamiltonian $H$ with a TTN state $|\psi_{ \{w\} }\rangle$, we need to find the
local tensors of TTN with which the expectation value of the Hamiltonian $E(\{w\}) = \langle\psi_{\{w\}}|H|\psi_{\{w\}}\rangle$ is minimized.

Unfortunately, there doesn't exist an exact solution for this optimization. In practical simulation, we can replace
the optimization equation with:
\begin{equation}
	E(w_i)={\rm min}_{w_i}\langle\psi_{\{w\}}|H|\psi_{\{w\}}\rangle
	\label{equ4}
\end{equation}
where we fix all the isometries except $w_i$. With this approximation, the optimization of $w_i$ in Eq.~\eqref{equ4} is easier. We
can optimize $w_i$ one by one. The whole procedure can be described as follows. Firstly, we generate a set of
 initially isometries $(w_1,w_2,\dots, w_n)$ randomly. Then we optimize one of them ($w_1$, for example) and
 fix the others according to Eq.~\eqref{equ4}. Once $w_1$ is optimized, we move to another isometry $w_2$ and optimize
 it in the same way. We iteratively perform this procedure until all the isometries are optimized. This procedure defines one sweep. Then we carry out the next sweeps until the energy is converged according to certain criteria. 

\begin{figure}[t]
	\includegraphics[width=80mm]{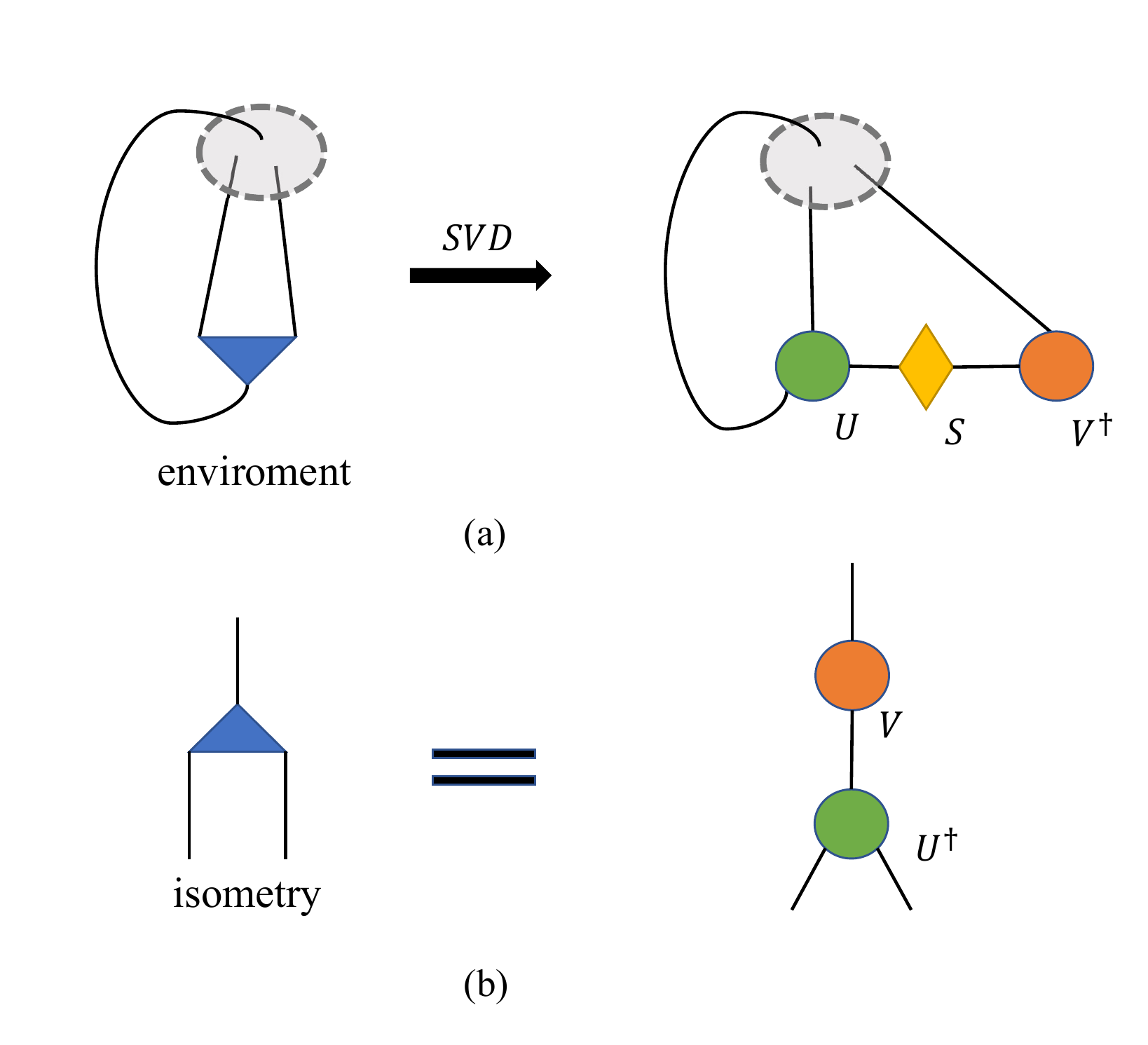}
	\caption{The optimization of the isometry. 
		After the environment tensor $\mathnormal{Y}$ is obtained,
		a SVD decomposition of it is performed $\mathnormal{Y}=USV^{\dagger}$.
		The optimized $w_i$ is set as $w_i=-VU^{\dagger}$. This step also takes a computational cost of $O(D^4)$.}
	\label{iso_svd}
\end{figure}

So the optimization of the TTN is reduced to minimize the energy in Eq.~\eqref{equ4}. However, we can not
solve this problem exactly either.
We follow the approach in \cite{PhysRevB.80.235127} where $w_i$ and $w_i^*$ are set to be independent and the
optimization problem in Eq.~\eqref{equ4} is reduced to
\begin{equation}
	E(w_i)= {\rm min}_{w_i}\langle\psi_{\{w\}}|H|\psi_{\{w\}}\rangle={\rm min}_{w_i}{\rm Tr}(\mathnormal{Y}w_i)
	\label{equ5}
\end{equation}
where we denote the environment of $w_i$ in $E(w_i)$ as $\mathnormal{Y}$.
The optimization problem in Eq.~\eqref{equ5} can be solved exactly. With a singular value decomposition of the environment
 $\mathnormal{Y}$:
\begin{equation}
	\mathnormal{Y}=USV^{\dagger}
\label{USVd}
\end{equation} 
the solution of Eq.~\eqref{equ5} can be obtained as $w_i=VU^{\dagger}$. We can easily see that this step has a cost
of $O(D^4)$. {Fig.~\ref{iso_svd} shows the procedures of optimizing isometries.}

In the following we will show that we can compute the environment $\mathnormal{Y}$ of $w_i$ at a cost of $O(D^4)$. For a
given local Hamiltonian, we can decompose it into a sum of a series of two-body operators. So
Eq.~\eqref{equ5} can be written as:
\begin{equation}
	E(w_i)={\rm min}_{w_i}\langle\psi|H|\psi\rangle
	={\rm min}_{w_i}\sum_{j,k}\langle\psi|H_{j,k}|\psi\rangle
\end{equation}
where $j,k$ denote the position of the two-body operator.

\begin{figure}[t]
	\includegraphics[width=80mm]{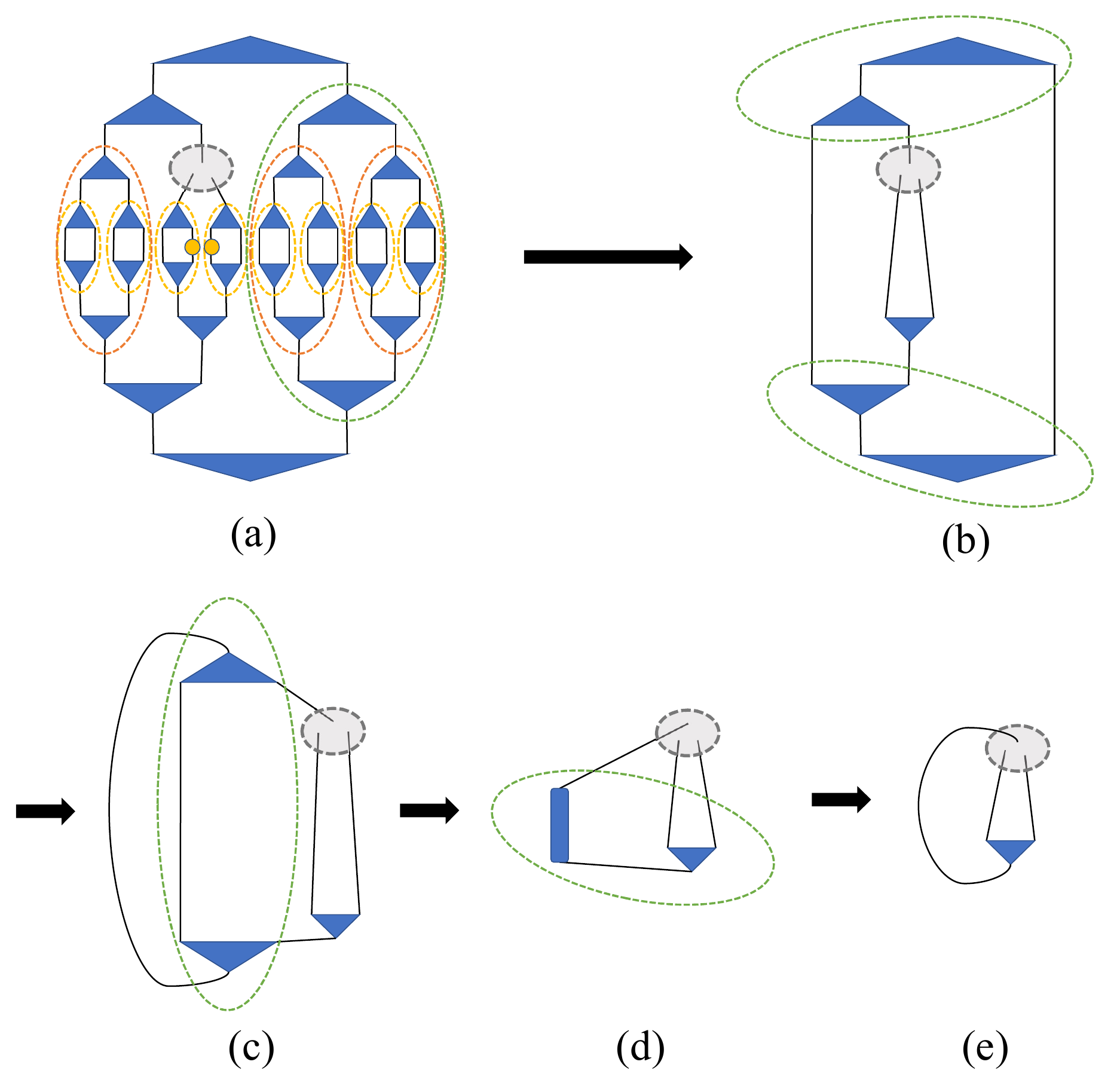}
	\caption{The calculation of the environment for an isometry. We can see that every step has a computational cost of $O(D^4)$. And the contractions which are not connected with the Hamiltonian term (denoted as the yellow circles)
		are just identities because of Eq.~\eqref{cons_iso}.}
	\label{opt_iso}
\end{figure}

Therefore, without loss of generality, we only need to deal with two-site operator (one site term is easy to handle as discussed below). We can further decompose the
two-site operator into a sum of direct product of two one-site operators whose number is $d^2$ in the worst case with $d$ 
the dimension of the physical degree of freedom at the physical  layer of TTN. Once we compute the contributions
 from all the terms in the Hamiltonian, we simply add those contributions to get the environment $\mathnormal{Y}$.
  As shown in Fig.~\ref{opt_iso}, the calculation of the environment $\mathnormal{Y}$ of $w_i$ for a two-site operator
  has a cost of $O(D^4)$. We can take advantage of Eq.~\eqref{cons_iso} to simplify the calculation as
  the contraction of two conjugated isometries results in an identity tensor if there is no operator acting on them.
  As the singular value decomposition of $\mathnormal{Y}$ has a cost of $O(D^4)$, the overall cost to
  optimize a TTN is $O(D^4)$. It is easily to show that the calculation of physical observables also have a cost of
  $O(D^4)$. So it is efficient to approximate the ground state of a local quantum Hamiltonian with a TTN.
  But as mentioned earlier, TTN can't capture the entanglement-entropic area law for systems in spatial dimension larger than two
  which hamper its application in simulating two-dimensional many-body systems.
   % Thus, the TTN structure is very efficient in computational cost, but it can not capture the entanglement-entropic area law for high-dimensional systems, and calls for the support from disentanglers in the MERA language(See Ref. \cite{PhysRevB.79.144108} for the details about disentanglers).

\section{Augmented Tree Tensor Network}
\label{sec_ATTN}

ATTN \cite{PhysRevLett.126.170603} was proposed to provide more entanglement entropy than TTN. 
%It can capture an area-law like entropy
%as $2$log$(L)$ and at the same time keep an algorithm complexity similar as TTN. As described in
%\cite{PhysRevLett.126.170603}, 
By placing disentanglers which are unitary tensors satisfying the following condition:
\begin{equation}
	\sum_{\beta_1\beta_2}(u_{\beta_1\beta_2}^{\alpha_1\alpha_2})(u_{\beta_1\beta_2}^{\alpha_1\alpha_2})^*=\delta_{{\alpha_1\alpha_2},({\alpha_1\alpha_2})^*}
	\label{cons_uni}
\end{equation}
at the physical  layer of TTN, the entanglement entropy encode in ATTN scales as $L\log(d)$ \cite{PhysRevLett.126.170603}. 
In ATTN, no couple of disentanglers are directly connected by an interaction term in the Hamiltonian to keep 
a low computational cost. ATTN has a computational complexity of $O(D^4d^2)$ which has the same scaling of $D$ as the TTN. However, a large improvement in numerical precision
of ATTN over TTN was shown in \cite{PhysRevLett.126.170603}.
%So, it proves that adding disentanglers is a very efficient way to augment the TTN, capturing the entanglement-entropic area law and keeping
% the computational complexity compatible with the TTN simultaneously.

\begin{figure}[t]
	\includegraphics[width=60mm]{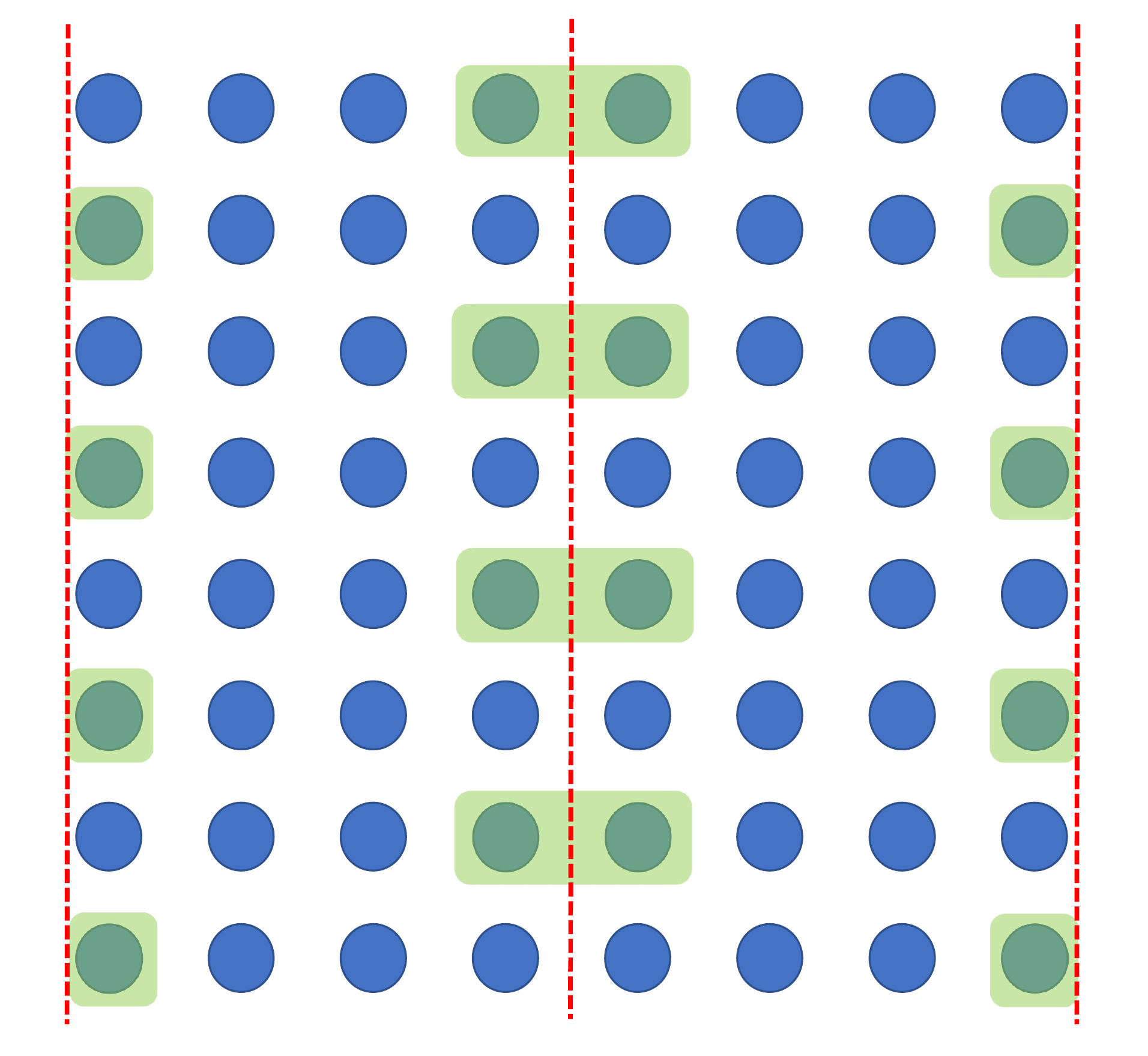}
	\caption{The strategy to place disentanglers in ATTN.  {Periodic boundary conditions are applied. Blue dots represent the physical layer of 2D TTN and green rectangles are disentanglers. } In order to maintain the same $O(D^4)$ scaling of computational cost as in TTN, no couple of disentanglers are directly connected by a Hamiltonian term. 
	As will be discussed late in the main text, this is not a necessary constraint to keep the $O(D^4)$ cost.}
	\label{ATTNdis}
\end{figure}

In Fig.~\ref{ATTNdis} we show the strategies of placing the disentanglers in ATTN for a $8\times 8$ lattice. We can see that a relatively small number of
 disentanglers are placed to encode the entanglement-entropic area law under the condition that no couple of disentanglers share the same Hamiltonian term.  
 %So, it's genera to find some improvements based on ATTN to get a much higher accuracy in numerical precision while keeping the computational complexity compatible
 %with ATTN by placing more disentanglers.

\section{Fully-Augmented Tree Tensor Network}
\label{sec_FATTN}
In this section, we propose another TTN based tensor network states which can provide more entanglement over ATTN and at the same time maintain the $O(D^4)$ computational cost.

\subsection{Position for disentanglers}
In ATTN, to maintain a computational complexity compatible with TTN, there are restrictions on how to place disentanglers at the physical layer. 
We find that the restriction that no couple of disentanglers share the same Hamiltonian term is not necessary to maintain a low computational cost.
%In ATTN, the restrictions are redundant and not necessary. 
%As we mentioned before, ATTN needs that no couple of disentanglers share the same Hamiltonian term.

%\begin{figure}[t]
%	\includegraphics[width=63mm]{NewTTN_dis}
%	\caption{An example to place the disentanglers in FATTN where no two disentanglers share the same physical index. There exist disentanglers across the boundary because
%	of the periodic boundary conditions.}
%	\label{NewTTN_dis}
%\end{figure}

\begin{figure}[t]
	\includegraphics[width=80mm]{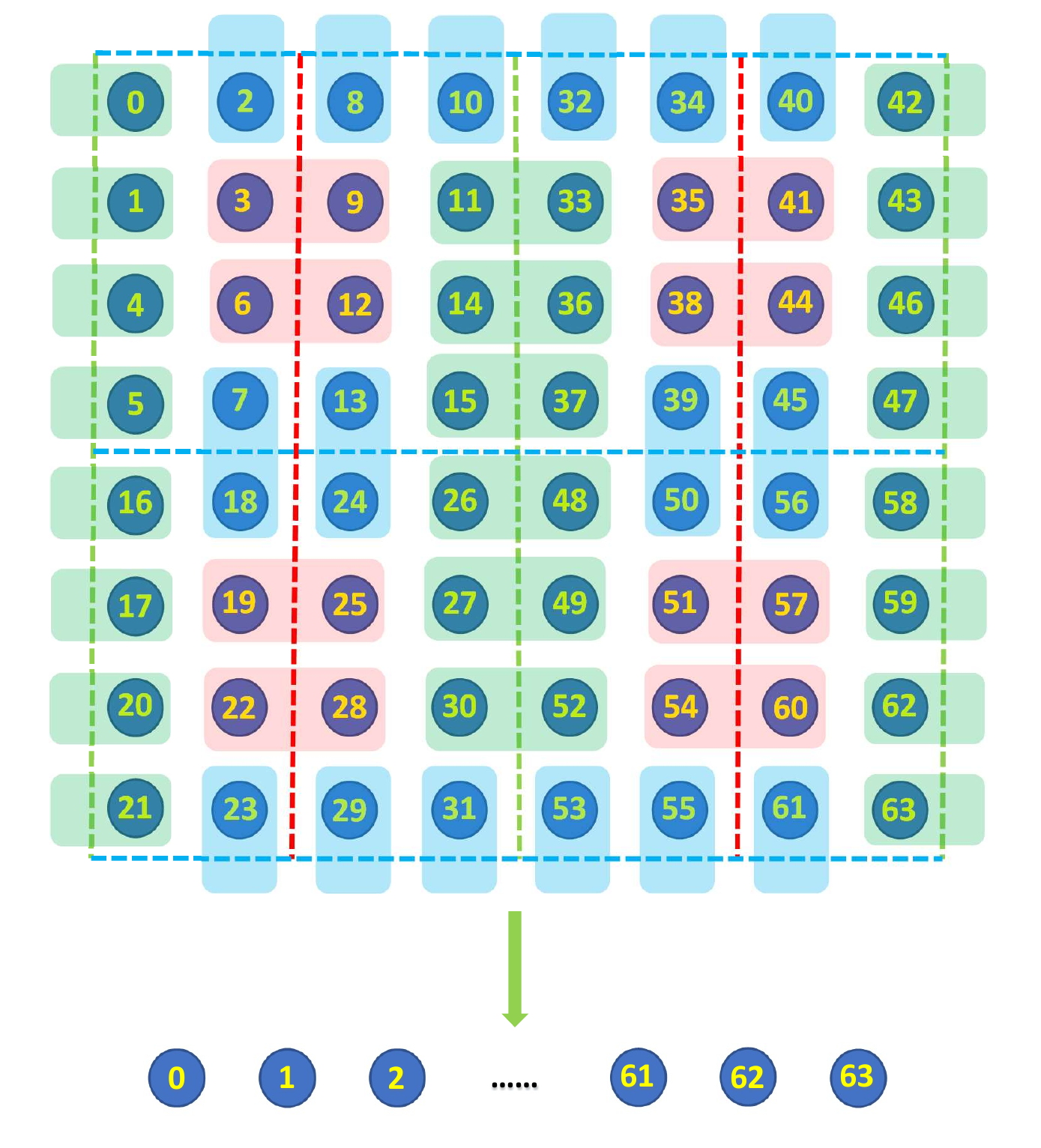}
	\caption{Strategy to put disentanglers in FATTN for a $8 \times 8$ lattice. 
		The shadowed bond represents a disentangler. Disentangler with different colors 
		are placed to compensate the violation of the entanglement entropic area law for the
		cuts with the same colors. And we use numbers to label the physical sites in 2D TTN in a 1D manner. The colored dash lines are the examples of cuts where the 2D TTN can not efficiently capture the entanglement-entropic area-law.
		See main text for more discussion.}
	\label{NewTTN_dis2}
\end{figure}

%We find that it is not essential to keep the complexity. Actually, we can place more disentanglers to encode the entanglement-entropic area law and still keep the computational
%cost at very low level.
We propose a new TTN based tensor network dubbed as Fully Augmented Tree Tensor Network (FATTN). In FATTN, we place disentanglers in a way that
no couple of disentanglers share the same physical index.
As we will discuss later, the computational cost in FATTN is $O(D^4d^4)$ with $D (d)$ the dimension of bond (physical) indexes. 
%and achieve a algorithm complexity of $O(D^4d^4)$.
So for an $L\times L$ lattice, we can add $L^2/2$ disentanglers at the physical layer.
We notice that if more disentanglers are added to the physical layer the TTN becomes a PEPS like
structure and exact contraction of them is infeasible in the calculation of physical observables. In Fig.~\ref{NewTTN_dis2}, we show an example on how to
place disentanglers on a $8\times 8$ lattice in FATTN. Given this constraint, we need to find a strategy to generate the maximum entanglement entropy using these
$8^2/2$ disentanglers for the $8\times8$ lattice.

%In order to find an optimized way to place disentanglers, we need to analyze where the TTN can not efficiently capture the entanglement-entropic area law. 

%For a 2D TTN, we can map it into an 1D TTN.
%We can label a 2D lattice  
%Let's consider a $8\times8$ lattice. 
As shown in Fig.~\ref{NewTTN_dis2}, We can number the physical indices at the physical  layer of the 2D TTN in a one dimensional manner. 
%Then, the 2D TTN trivially become a standard 1D TTN. The only different thing is that some hamiltonian terms aren't local as we may encounter
%some long range interactions seeing from the 1D TTN. However, 
Using this labeling scheme, we can easily find the cut where the TTN can not capture the entanglement-entropic area law.
For a small bond-dimension $D$, examples for cuts
where the 2D TTN can not capture the entanglement-entropic area law are shown in the Fig.~\ref{NewTTN_dis2} by colored dash lines.
%(For larger systems as $16\times 16, 32\times 32 \cdots$, we can iteratively use Fig.~\ref{TTN_arealaw} to find where the
%TTN can not efficiently encode the entanglement-entropic area law)
So when augmenting the TTN with disentanglers, we mainly consider these cuts. For concreteness, only periodic boundary conditions are considered in this work.
%(We should mention that we consider the periodic boundary conditions throughout the paper. For other boundary conditions, it is quite easy to adapt to).

We firstly consider cuts where the system is divided into two parts with equal number of sites. Fig.~\ref{NewTTN_dis2} shows three ways to bipartite $\mathnormal{A}$ and
$\mathnormal{B}$. For the cut with green (blue, red) line, suppose we put $n_g (n_b, n_r)$ disentanglers cross them, the entanglement entropy along these cuts are:
\begin{equation}
\begin{cases}
	S_g = \log(I_g)=\log(d^{2n_g}D)\\ 
	S_b = \log(I_b)=\log(d^{2n_b}D^2)\\
	S_r = \log(I_r)=\log(d^{2n_r}D^4)
	\end{cases}
\end{equation}
%The entropy assessed by the blue line is:
%\begin{equation}
%	S_b=\log(I_b)=\log(d^{2n_b}D^2)
%\end{equation}
%if we put $n_b$ disentanglers along the red line. Similarly, the entropy along the red line is:
%\begin{equation}
%	S_r=\log(I_r)=\log(d^{2n_r}D^4)
%\end{equation} 
As will be discussed below, a constraint that no couple of disentanglers share the same physical index needs to be satisfied 
to maintain a low computational cost. We also need to keep $I_g\approx I_b\approx I_r$ to make a balance among different cuts to maximize the lowest entanglement entropy
with these cuts which means
\begin{equation}
	\begin{cases}
		n_g+n_b+n_r\leq 32\\
		n_g\geq n_b\geq n_r
	\end{cases}
\end{equation}

%Considering all the conditions mentioned above, 
We can set $n_g=14,n_b=10,n_r=8$ {(The choice of $n_g, n_b, n_r$ is not unique. In principle, the optimal choice of them changes with the increase of bond-dimension)} and the position for disentanglers is shown in Fig.~\ref{NewTTN_dis2}. For an equal bipartition of the $8 \times 8$ system, the length
of a cut is $16$. So to capture an area law like entanglement entropy $S = 16 \log k = \log(k^{16})$, the bond-dimensions required by different bipartitions are:
\begin{equation}
	\begin{cases}
		D_g=d^4(\frac{k}{d^2})^{16}\\
		D_b=d^6(\frac{k}{d^2})^8\\
		D_r=d^4(\frac{k}{d^2})^4\\
	\end{cases}
	\label{bdm}
\end{equation}
From Eq.~\eqref{bdm} we can see that for $k \le d^2$, with a relatively small bond-dimension, FATTN can capture an area law like entanglement entropy $L\log(d^2)$.
For most of other bipartitions, the entanglement entropy provided in FATTN are also $L\log(d^2)$ with a similar requirement on $D$ (see the discussion in the Appendix). 
Comparing with ATTN, the entanglement entropy in FATTN is doubled from $L\log(d)$ to  $L\log(d^2)$ on a finite lattice for most of the cuts with a large $D$.
At the same time the computational cost is $O(D^4d^4)$ which is comparable to $O(D^4d^2)$ for ATTN as will be discussed in the next subsection.

%And in other words, we want to use a relatively small bond-dimension $D$ to reproduce an entropy of $L\log(4)$. Actually, the ATTN can capture the entanglement-entropic area law, but suffers from the limitation of the maximum entropy is $L\log(2)$. We break through this limitation by placing more disentanglers. More importantly, we need to point out that the computational cost of the FATTN is compatible with ATTN, which is $O(D^4d^4)$.

%Another crucial point is that we can regain the locality for the hamiltonian in the 2D TTN by adding disentanglers at the bottom layer. For example, we consider the interaction between site $26$ and site $48$ shown in the Fig.~\ref{NewTTN_dis2}. It's a long range interaction seeing from the 1D TTN. However, through the extra disentanglers, we can regain the locality for the interaction term. And in reference Ref. \cite{cataldi2021hilbert}, it shows that locality is very important to get a higher accuracy. So, adding disentanglers is a fully staffed approach to augment TTN.

\subsection{Optimization of FATTN}

\begin{figure}[t]
	\includegraphics[width=80mm]{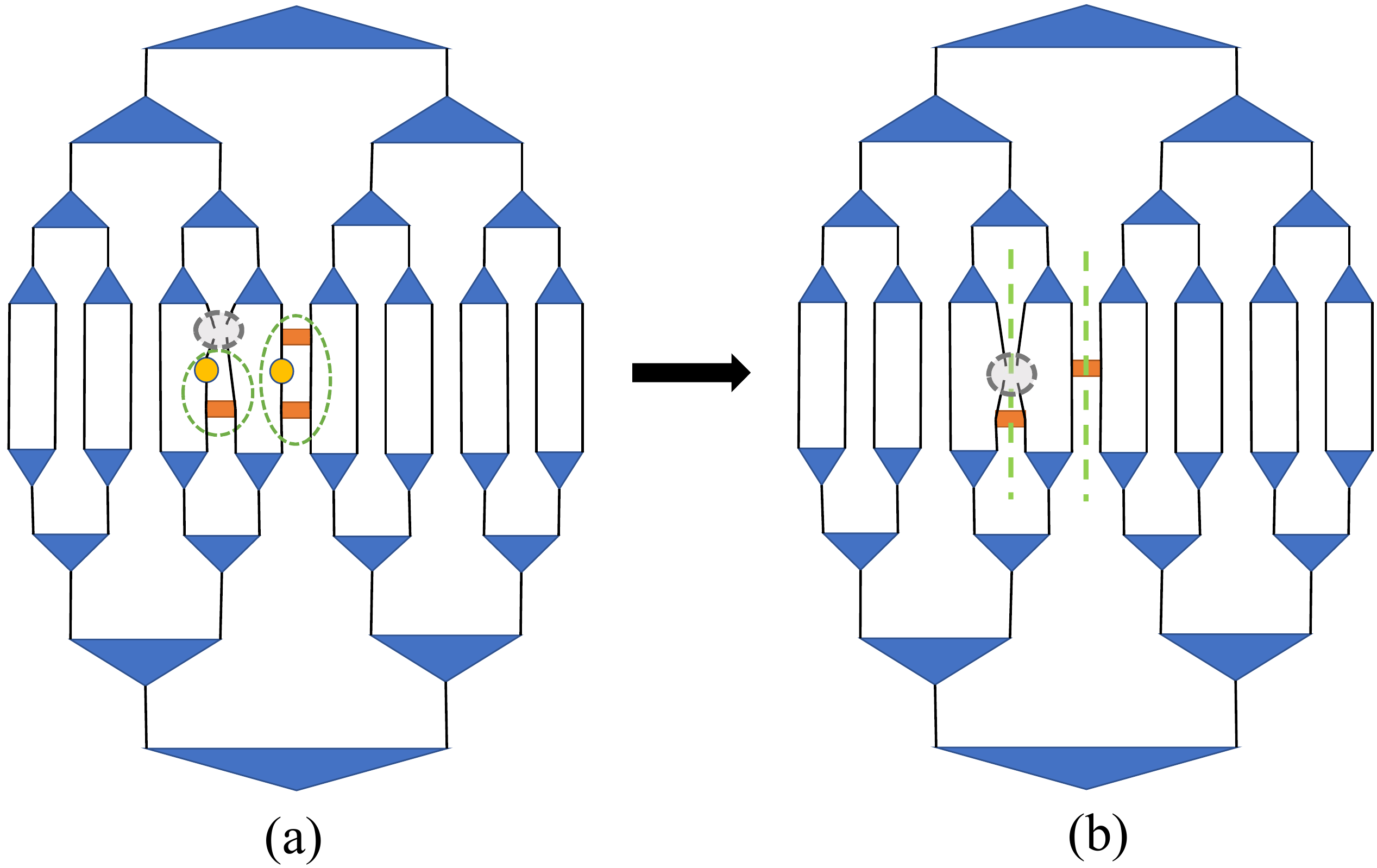}
	\caption{In the optimization of disentangler, FATTN can be reduced to a TTN like structure with an SVD. Here we only show the disentanglers which are connected with the considered Hamiltonian term.}
	\label{disentangler1}
\end{figure}

\begin{figure}[t]
	\includegraphics[width=80mm]{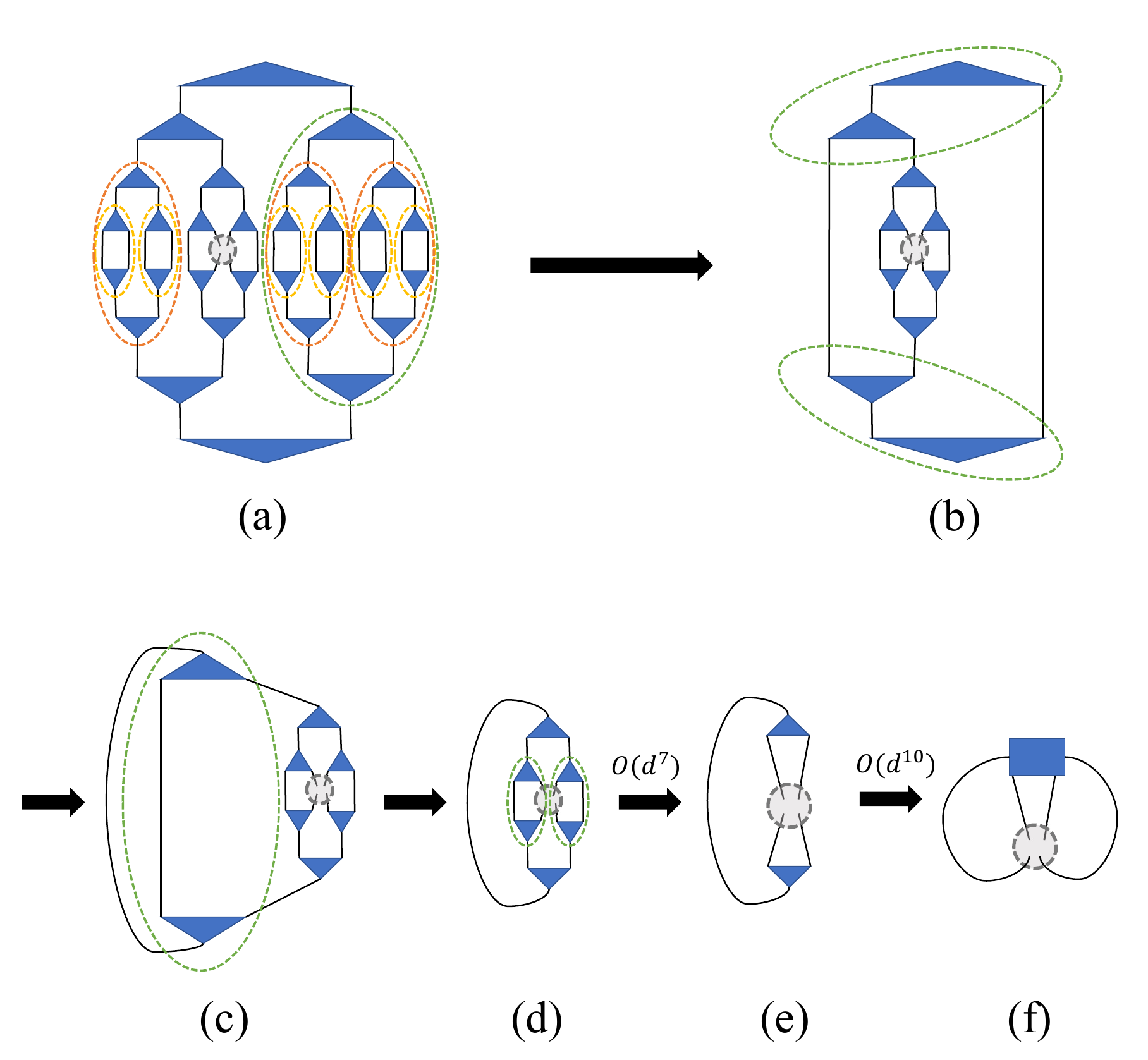}
	\caption{The computation of the environment for disentanglers. The cost of this step is $O(D^4)$.}
	\label{opt_dis}
\end{figure}

\begin{figure}[t]
	\includegraphics[width=80mm]{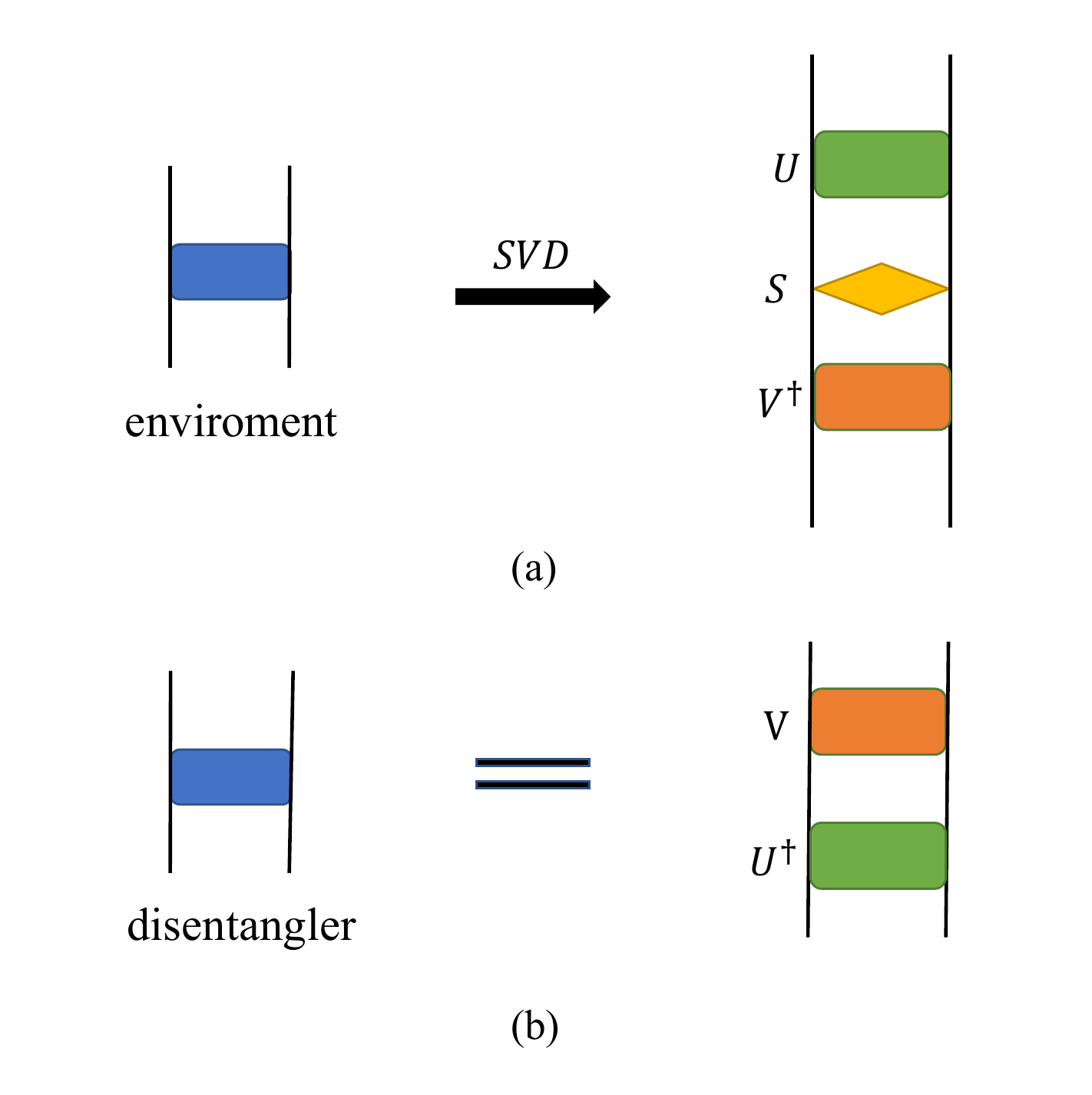}
	\caption{The optimization of disentangler from the corresponding environment tensor, similar as in Eq.~\eqref{USVd}.}
	\label{disnew}
\end{figure}

To optimize a FATTN, we need to optimize both the disentanglers and the isometries in it.
We follow the procedure in the MERA \cite{PhysRevB.79.144108} to optimize disentangler which is very similar to the optimization of isometry in the TTN.
 First, we calculate the environment for the disentangler we want to optimize. For a Hamiltonian term, we just need to consider the disentanglers which are
 directly connected to it because other disentanglers can be annihilated because of Eq.~\eqref{cons_uni}.
As show in Fig.~\ref{disentangler1}, we firstly take a singular value decomposition to get a standard TTN. Then, we contract the TTN
 to get the environment tensor. This procedure takes a cost of $O(D^4)$ as shown in Fig.~\ref{opt_dis}. Once we obtain the environment tensor, we follow the strategy
 in the TTN to get the optimized disentanglers. The details can be found in Fig.~\ref{disnew}.

\begin{figure}[t]
	\includegraphics[width=80mm]{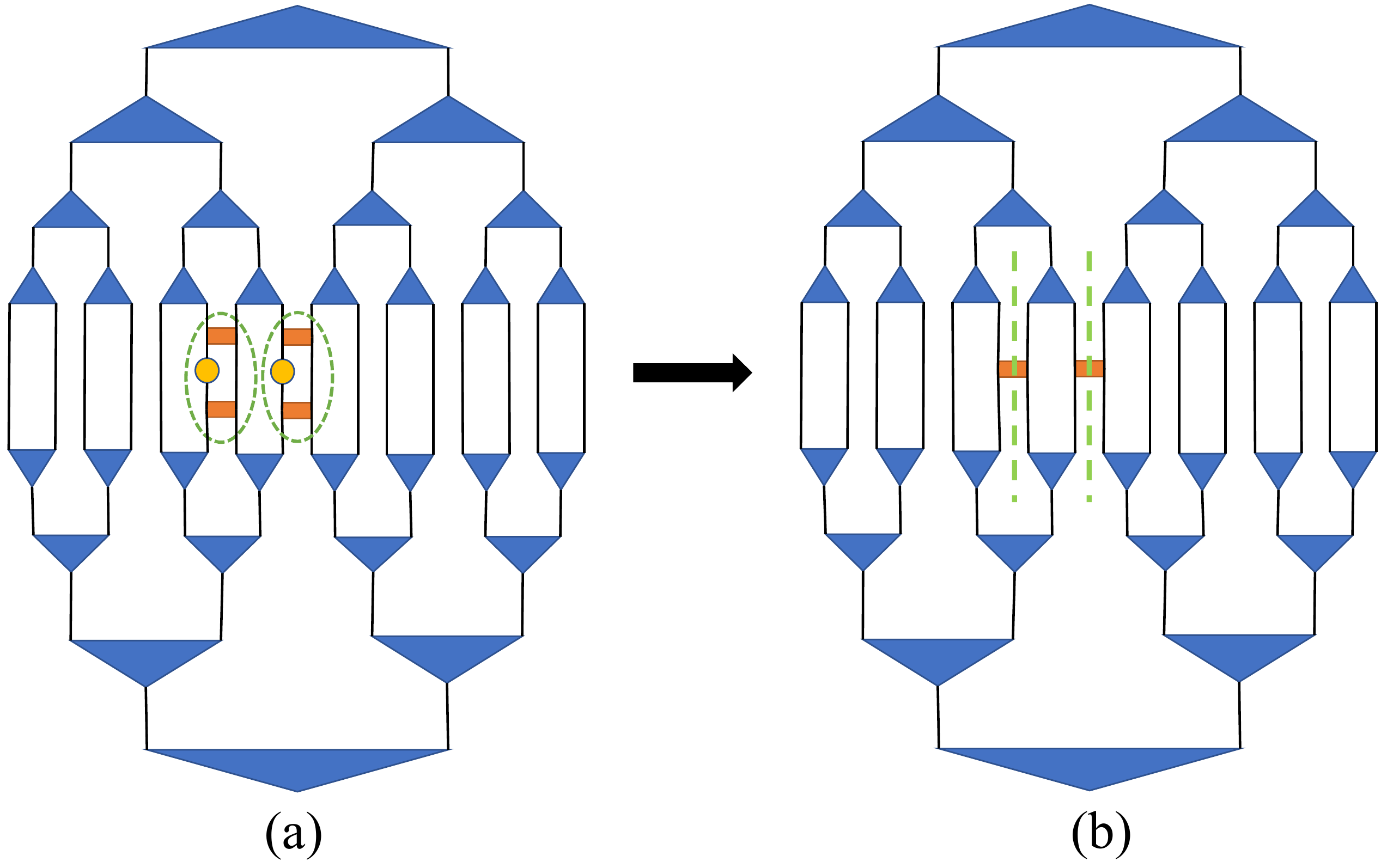}
	\caption{Similar as in Fig.~\ref{disentangler1}, in the optimization of isometry, FATTN can be reduced to a TTN like structure with an SVD. Here we only show the disentanglers which are connected with the considered Hamiltonian term.}
	\label{iso_dec}
\end{figure}

The procedure to optimize isometries is similar. For a Hamiltonian term, we only need to consider the disentanglers which are directly connected to it. We contract
the Hamiltonian term and the connected disentanglers to get an effective Hamiltonian term in same spirit of ATTN \cite{PhysRevLett.126.170603}. We then decompose the partial contracted FATTN into a sum of a series of standard TTN as shown in the Fig.~\ref{iso_dec}. After this, we can follow the procedure in the TTN to optimize the isometries in the FATTN.
%\C{check this part}

Overall, for an arbitrary Hamiltonian term, we decompose the relevant disentanglers (for a two-site operator, there are two disentanglers need to be considered in the worst
case which
causes a factor in the computational cost of $d^4$). Then, FATTN  becomes a standard TTN. So the overall computational complexity is $O(D^4d^4)$. 
%The only drawback of the FATTN compared
%with the ATTN is that we can not optimize disentanglers parallel.
%And this may increase some computational cost, but it is not the point as we will see the FATTN is the most efficient method to get a higher accuracy with the same computational cost in our actual simulations.

\section{Benchmark results}
\label{sec_benck}

To show the accuracy of the FATTN, we calculate the ground state energy of the transverse Ising model \cite{Suzuki2013} with the Hamiltonian
\begin{equation}
H_{\rm Ising}=-\sum_{\langle i,j \rangle}\sigma_{i}^{x}\sigma_{j}^{x}-\lambda\sum_{i}\sigma_{i}^{z}
\end{equation}
where $\sigma_x$ and $\sigma_z$ are Pauli matrices and $\lambda$ is the strength of the transverse magnetic field. We consider a $8\times8$ lattice with periodic boundary conditions
where highly accurate numerical results for energy are available for benchmark \cite{PhysRevB.80.235127}.
We focus on  $\lambda=3.05$ close to the critical point which is the hard region of this model.
%(For lager systems, we can iteratively use strategies shown in the Fig.~\ref{NewTTN_dis2} to get systems size like, $16\times 16$,$32\times 32 \cdots$. for the position of disentanglers. In this paper, we just give a simple calculation of the $16\times 16$ lattice size for lager systems).
%We know that the model undergoes a quantum phase transition for $\lambda \approx 3.044$ in the thermodynamic limit \cite{refId0,PhysRevE.66.066110}. And for our $8\times8$ lattice, we set $\lambda=3.05$ as the value of the transverse magnetic field at the critical point. We compare the estimated ground state energy with the best known results obtained via TTN \cite{PhysRevB.80.235127} at the critical point.
%We also consider $\lambda=2,4$ to benchmark our results, and compare our results with the data calculated by DMRG. 

\begin{figure}[t]
	\includegraphics[width=80mm]{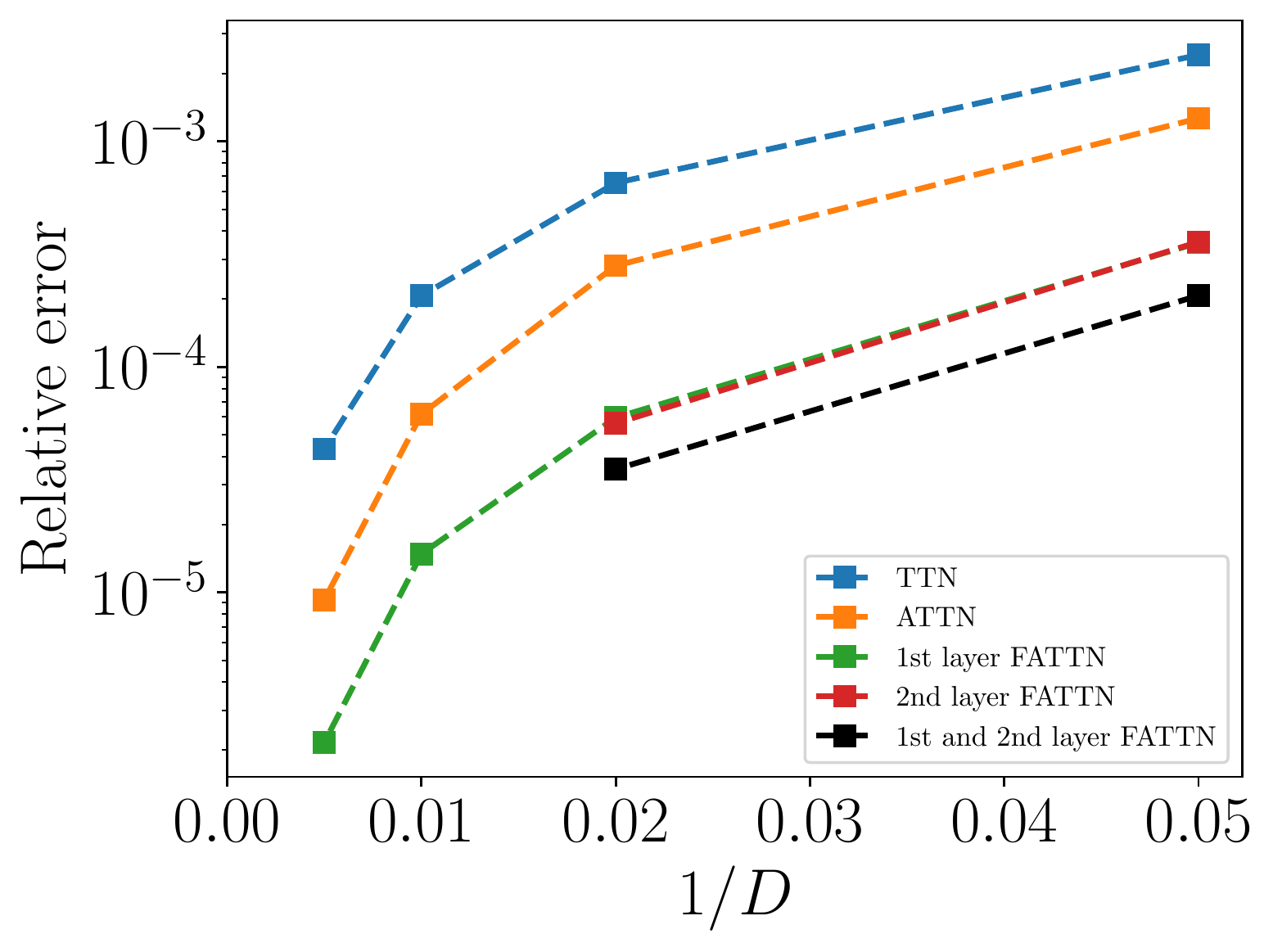}
	\caption{Relative error of the ground state energy  {per site} for different tensor network states ansatzes of the transverse Ising model near the critical point ($\lambda = 3.05$) for a $8\times 8$ lattice.}
	\label{mul_result}
\end{figure}

In Fig.~\ref{mul_result}, we show the relative error of the ground state energy per site as a function of the bond dimension at $\lambda=3.05$ for TTN, ATTN, and FATTN.
As expected, FATTN gives the lowest energy with fixed $D$. The error with FATTN is reduced by one (a half) order of magnitude over TTN (ATTN) for all bond dimensions.

 %And the most crucial point is that FATTN is the most efficient method to get a higher accuracy among the methods mentioned above with the same computational cost.

%And more importantly, we see that the energy converges roughly the same with bond dimension for the TTN, ATTN and FATTN. This finding is genera, since the entropy
% those tensor network structures can encode has the same relation with the bond dimension $D$, scaling as $cL+log(D)$. 

\begin{figure}[t]
	\includegraphics[width=80mm]{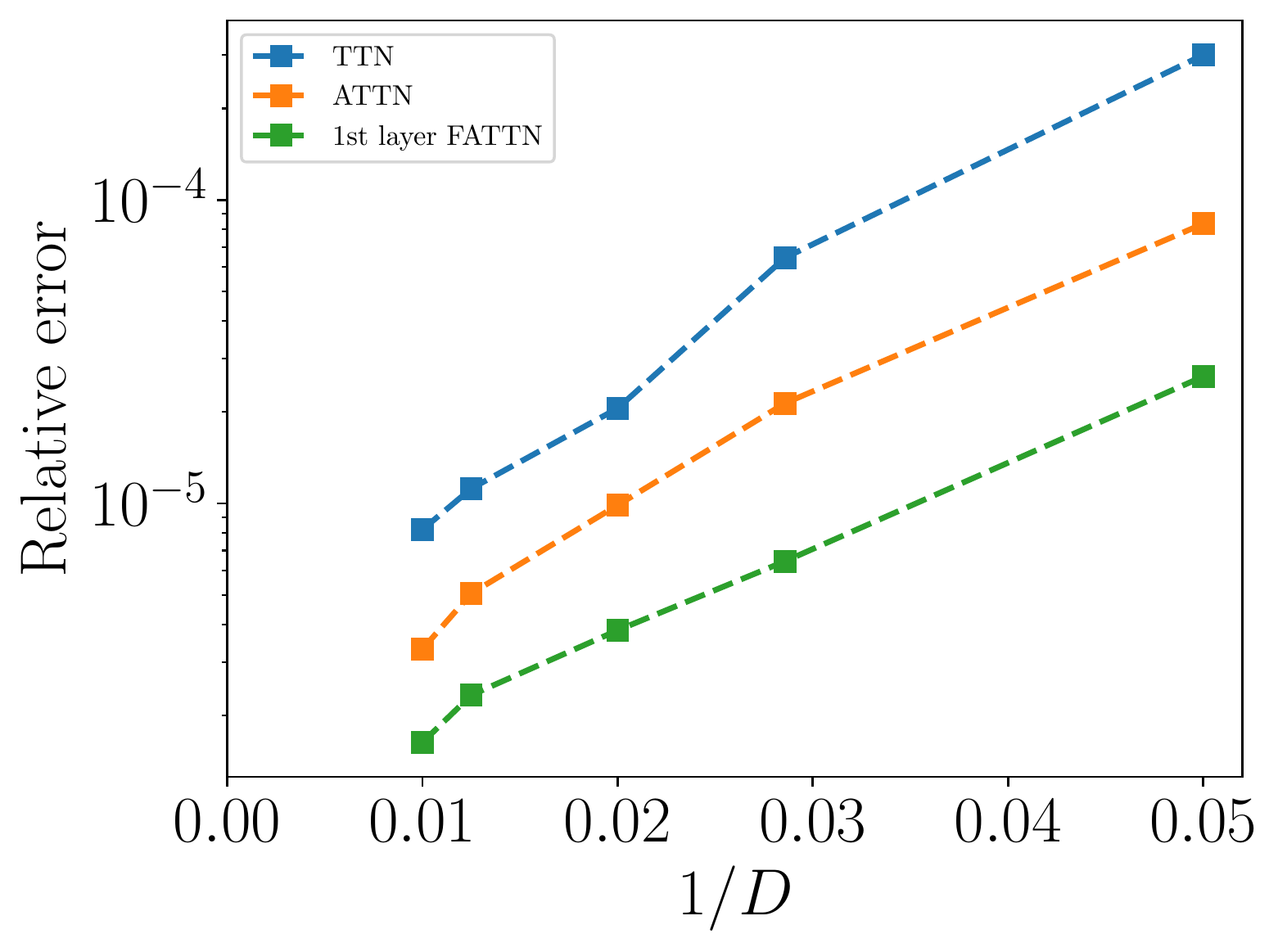}
	\includegraphics[width=80mm]{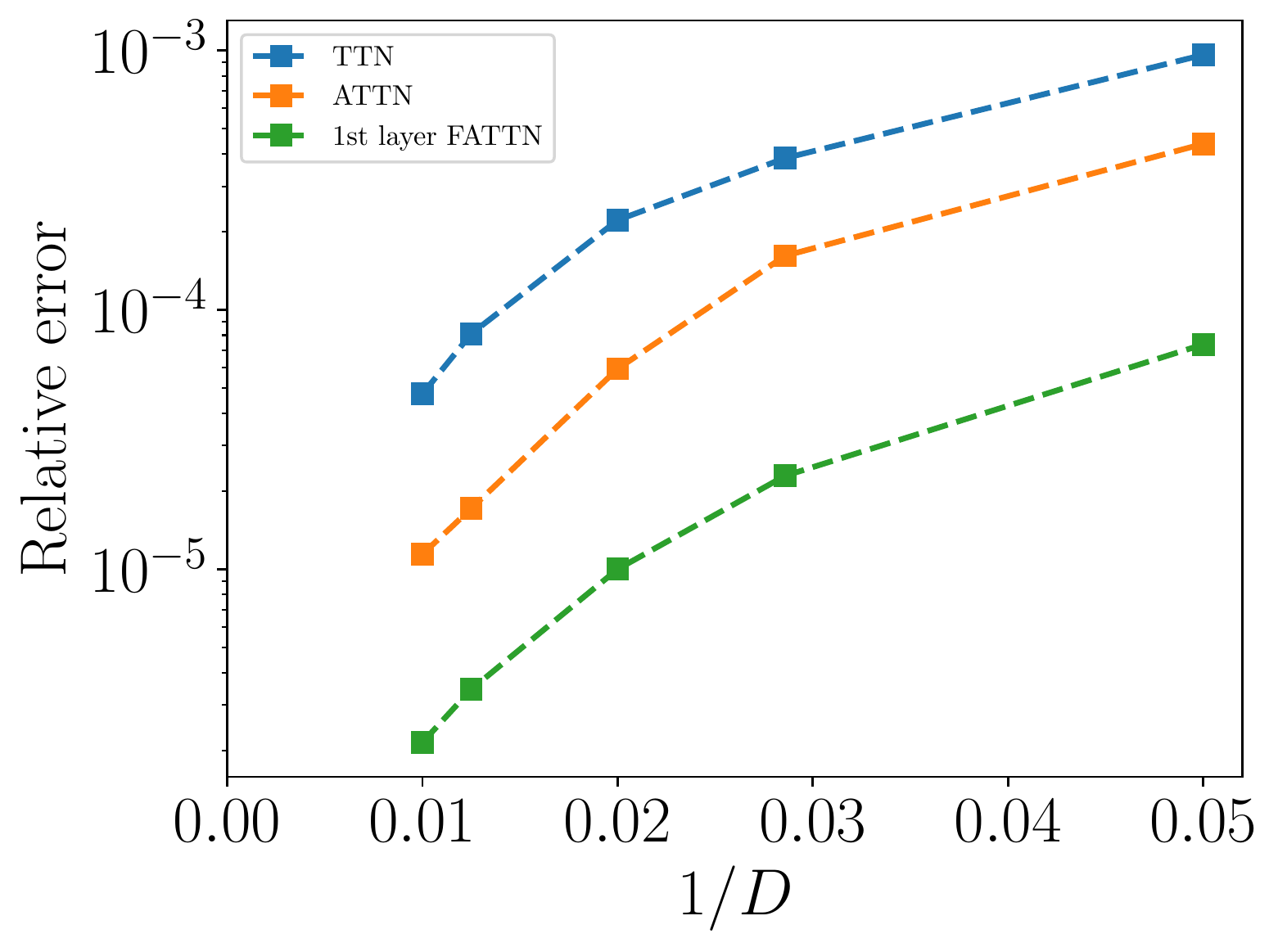}
	\caption{Relative error of the ground state energy {per site} as a function of bond dimension for different tensor network states ansatzes away from the critical point for a $8\times 8$ lattice. Upper: $\lambda=2$. Down: $\lambda=4$.
	}
	\label{result24}
\end{figure}

The results away from the critical point are shown in Fig.~\ref{result24}. We can see a similar improvement of FATTN over TTN and ATTN for $\lambda = 2$ and $4$.
%Away from the critical point, the results compare very well as shown in Fig.~\ref{result24}, however not as well as the critical point. Actually, for $\lambda=2$, we
%find that the gap of the relative error between the TTN and the FATTN becomes smaller and smaller with the increasing of the bond-dimensional. But, we can see that the
%FATTN is still much better than the TTN and the ATTN.

%And for $\lambda=4$, the scaling behavior of the relative error with bond-dimension is the same with $\lambda=3.05$. This is expected as the spectrum of the entropy at $\lambda\geq4$ is roughly the same with $\lambda=3.05$. And we can see an improvement roughly the same with the simulations of $\lambda=3.05$. But, actually, for the $\lambda\geq4$, we get a higher numerical precision than $\lambda=3.05$, which corresponds to the critical behavior of the transverse Ising model.

%\begin{figure}[t]
%	\includegraphics[width=80mm]{transition}
%	\caption{Expectation value of the transverse magnetization as a function of the transverse magnetic field for a $8\times 8$ lattice.}
%	\label{transition}
%\end{figure}

\begin{figure}[t]
	\includegraphics[width=80mm]{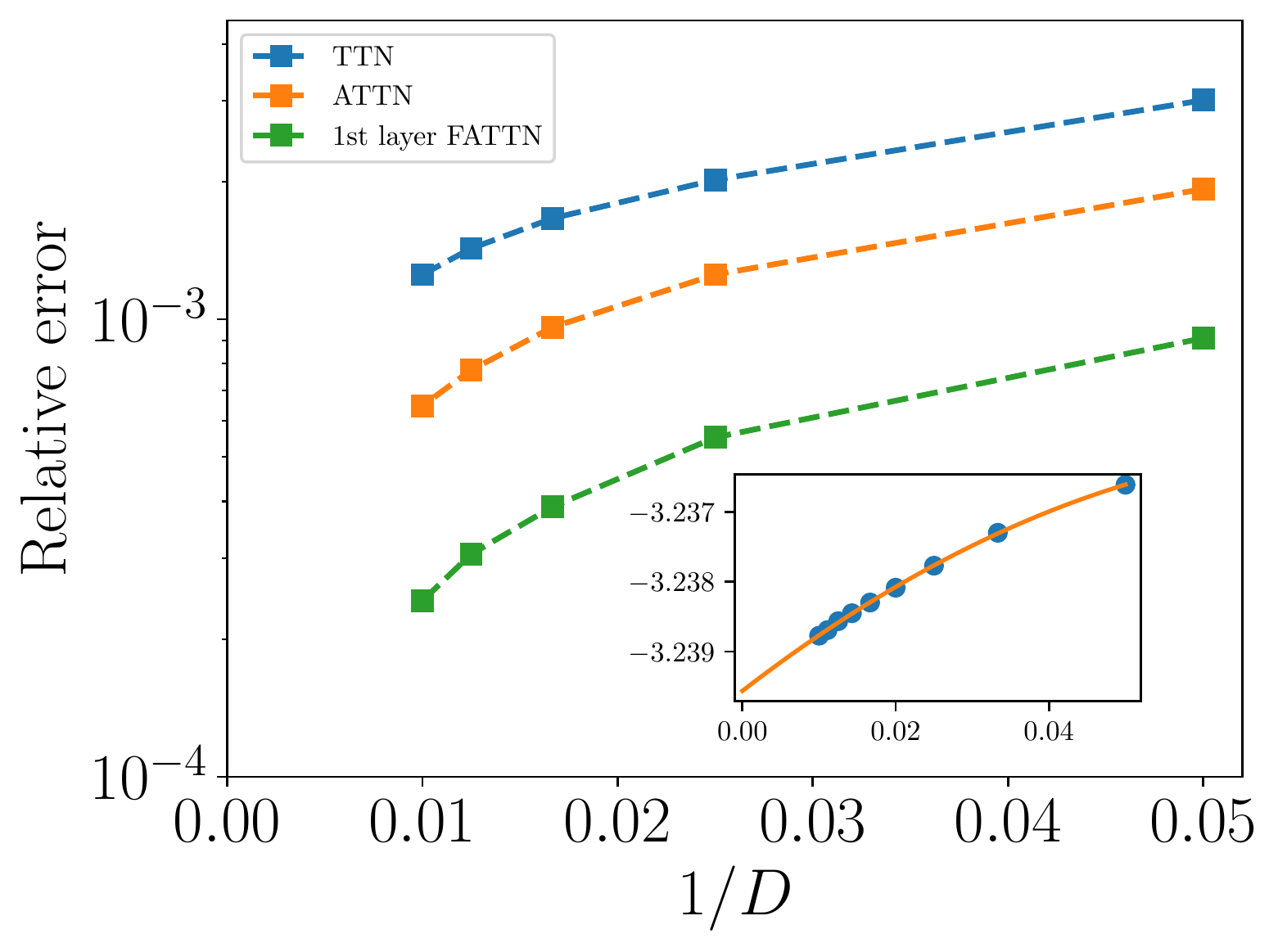}
	\caption{Relative error of the ground state energy per site for different tensor network states ansatzes of the transverse Ising model near the critical point ($\lambda = 3.05$) for a $16\times 16$ lattice, compared with the result extrapolated from FATTN data. {
			In the inset, we show the quadratic fit of the FATTN energy and the extrapolated value is $-3.23956(2)$.} }
	\label{results16}  
\end{figure}

%In Fig.~\ref{transition}, we show the transverse magnetization per site in the ground state of the $8 \times 8$ transverse Ising model. 
%From the Fig.~\ref{transition}, we can see a stiff dependence on $\lambda$ around $\lambda=3$ which is consistent with the presence of a phase transition at $\lambda=3.044$ in
%the thermodynamic limit. 
For lager systems, we do not have accurate results to compare against. In Fig.~\ref{results16}, we calculate the relative error $\Delta E=|(E_{\rm D}-E_{\rm extra})/E_{\rm extra}|$ of the ground state energy per site for different tensor network ansatzes at different bond-dimension $D$ for a $16 \times 16$ lattice. We extrapolate the results calculated by FATTN to obtain $E_{\rm extra}$. We also find a significant improvement in the simulations of FATTN over TTN and ATTN.
{We notice that the $16\times 16$ FATTN energy with $D=20$ is lower than TTN with $D = 100$ while the $8\times 8$ FATTN energy with $D=20$ is comparable with TTN $D=100$. }

{All of the computations are performed on a single node  which contains 2 Intel Xeon Gold 6420R CPUs. For $D=100$, $8\times 8$ lattice case, TTN consumes 25 seconds per sweep while ATTN consumes 60 seconds and FATTN consumes 209 seconds per sweep, which is compatible with the scaling analysis. All the calculations above need about 1 GB RAM per process.}

\section{Interpolation between TTN and MERA}
\label{sec_multi}
In the above section, we only place disentanglers at the physical  layer of TTN. But this is not the only scheme to augment a TTN. We can place disentanglers at
any layer or even at multiple layers of the TTN. The key issue is to keep the computational complexity affordable. 
%We notice that if disentanglers
%are placed in every layer of TTN, the TTN becomes a MERA \cite{PhysRevB.79.144108} which is highly costly.

\subsection{Single-layer FATTN}
The bond dimension of the disentangler depends on its position in the TTN which determines the upper bound on the entanglement entropy the FATTN can capture.
If disentanglers are placed at the physical (first) layer of TTN, the bond dimension of disentangler is $d$ ($d = 2$ for spin $1/2$ system) and the entropy the FATTN can mostly encode is
$S = L\log(d^2)$.

\begin{figure}[t]
	\includegraphics[width=80mm]{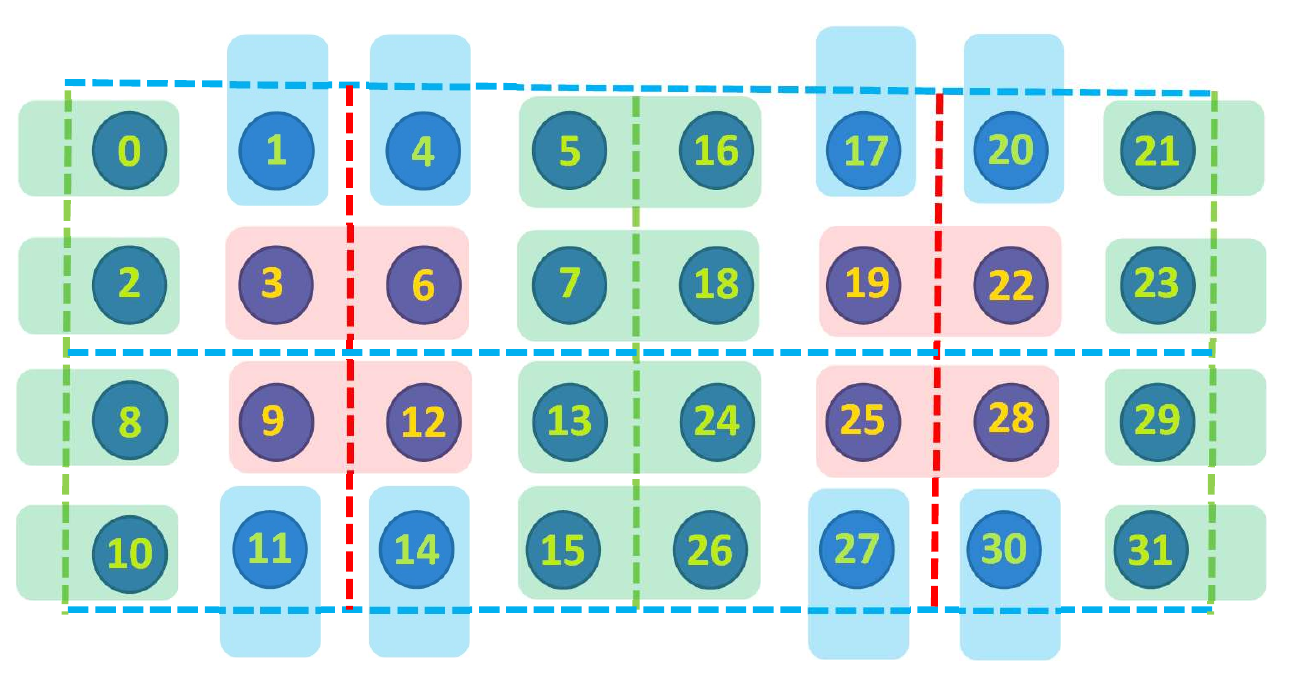}
	\caption{The strategy to place disentanglers in 2nd layer FATTN. Notice that no couple of disentanglers share the same site to ensure a low computational cost.}
	\label{second_dis}
\end{figure}

%It is clearly limited by the dimension of the disentanglers' indices at the bottom layer. 
If we put disentanglers in higher layer, the bond dimension of disentangler is larger than $d$
(to maintain a low cost, we also need to ensure no couple of disentanglers share the same site).
%A good strategy of the second-lowest layer is shown in Fig.~\ref{second_dis}).
But a balance between the bond dimension of disentangler and the maximum number of disentanglers can be placed
needs to be reached to provide the maximum entanglement entropy in FATTN. As shown in Fig.~\ref{second_dis}, we can only place $8^2/4$ disentanglers in the 
second layer of the TTN, but the bond dimension of the disentangler becomes $d^2$. So the entanglement entropy
can be encoded when placing disentanglers in second layer is the same as the first layer case.  

However, when placing disentanglers in higher layer, we find that the entanglement entropy is limited by the red line cuts shown in Fig.~\ref{second_dis} or Fig.~\ref{NewTTN_dis2}.
The entanglement entropy can provide is $S=L/4\log(D)$ if $D < d^8$ (for $D > d^8$, the entanglement entropy the FATTN can provide is still $L\log(d^2)$). 
%So in order to enlarge the entropy, we need to place multi-layer disentanglers in the TTN to achieve an acceptable computation cost.

\begin{table}[t]
	\begin{tabular}{ccc}
	    \toprule
	    Structure& Computational Complexity\\
	    \hline
	    1st layer & $O(D^4d^4)$\\
		2nd layer &  $O(D^4d^8)$\\
		3rd layer &  $O(D^4d^{16})$\\
		1st \& 2nd layer & $O(D^4d^{20})$\\
		ATTN& $O(D^4d^2)$\\
		\hline
		
	\end{tabular}
	\caption{The computational complexity for FATTN with different ways to augment the TTN with disentanglers.
		The results for ATTN are also listed for comparison. }
	\label{stu_table}  
\end{table}

In TABLE.~\ref{stu_table}, we list the computational complexity for different types of single-layer FATTN.
%The analysis of the computational complexity is just the same with the FATTN, where we firstly decompose the relevant disentanglers to get the standard TTN and then use the procedures in the TTN to optimize and contract our tensor network structure. We should mention that the analysis in the table.~\ref{stu_table} isn't rigorously right about the entropy and is a proximate analysis. Indeed, it depends on many factors e.g., the strategies of placing disentanglers,  the max bond-dimension $D_{max}$ and the lattice size(In our analysis, we just consider a relatively small bond-dimension $D$ to benchmark our results).

\subsection{Multi-layer FATTN}
We can also place disentanglers on multiple layers of the TTN. It is easily to show that the FATTN is actually a 2D MERA if disentanglers are placed on every layer of the TTN.
%This is slightly more complex than the previous task, as we may encounter some problems in algorithm complexity if we do not have a good strategy to position disentanglers. And in the worst case, our structure would become a 2D MERA.
%2D MERA can provide a real area-law entanglement entropy (see below )but it is highly costly with a computational complexity $O(D^{16})$ \cite{PhysRevLett.102.180406} which limits the bond dimension we can reached. 

%The entanglement entropy in 2D MERA can be written as: 
%\begin{equation}
%	S =L\log(d^2)+\cdots + L/2^n\log(D^2) \approx cL\log(D)
%\label{MERA_ENT}
%\end{equation}
%where each term in the summation is the contribution of a layer in MERA.
2D MERA can provide an area-law entanglement entropy proportional to $L\log(D)$. 
But it is highly costly with a computational complexity $O(D^{16})$ \cite{PhysRevLett.102.180406} which limits the bond dimension can be reached.

To make a balance between cost and entanglement entropy captured in the wave-function ansatz,
we can interpolate between single layer FATTN and MERA by placing disentanglers on a relatively small number of layers of the TTN.
%Considering this, if we want to increase the entropy of our single layer TTN, we can add more layers of disentanglers. But, we also want to keep the computational complexity at an acceptable range. So, we just put a relatively small number of disentangler layers in the TTN to prevent it from becoming 2D MERA, leading to a disaster in computational cost.
As an example, we place disentanglers
in both the first and second layer of TTN according to Fig.~\ref{NewTTN_dis2} and Fig.~\ref{second_dis}. The computational complexity are shown
in Table.~\ref{stu_table}. 
%What we want to point out here is that we can add less disentangler layers to reproduce the entropy the MERA can encode and get a lower computational cost. For the third-layer TTN shown in the Table.~\ref{stu_table}, the entropy is $L/4log(D)$ for a relatively small $D$ and the computational complexity is $O(D^4d^{16})$. So, we need a computational cost of $O(D^{16}d^{16})$, compatible with the MERA \cite{PhysRevB.79.144108} proposed by G. Vial, to encode an entropy of $Llog(D)$ that the MERA can capture. However, the structure of our Third-layer TTN is much simple and can be easily augmented by using more layers of disentanglers.
%More importantly, we can seek some other fancy ways to place multi-layer disentanglers to get a better performance than MERA both in computational cost and the numerical precision.

\subsection{Results for the single-layer FATTN and multi-layer FATTN}

In Fig.~\ref{mul_result}, we show the comparison of the relative error of the ground state energy for different tensor network ansatzes.
The system is a $8 \times 8$ transverse Ising model with $\lambda = 3.05$. The result of 2nd layer FATTN is comparable with 1st layer FATTN, but the 2nd layer FATTN has a higher cost.
When placing disentanglers on both the first and second layer, a tiny improvement can be achieved.
From Fig.~\ref{mul_result}, we can find that the best strategy for FATTN is to place the disentanglers at the physical layer according to Fig.~\ref{NewTTN_dis2}.
%This conclusion can also been drawn from Eq.~\eqref{MERA_ENT} where the contribution of entanglement entropy decays exponentially from bottom to top layers in MERA.

\section{Conclusions}
\label{sec_con}
In this work, we propose a new tensor network state ansatz FATTN by releasing the unnecessary constraint in ATTN. 
FATTN can provide an area law like entanglement entropy scaling as $L\log(d^2)$ for most of the cuts (with a large bond dimension $D$ on a finite lattice) with a low computational cost $O(D^4d^4)$ if the
disentanglers are placed on the physical layer. The benchmark results on 2D transverse Ising model near critical point
show a large improvement over TTN (nearly one order of magnitude) and ATTN (nearly a half order of magnitude) on the ground state energy. FATTN can be viewed
as an interpolation between TTN and MERA to reach a balance between the computational cost and entanglement entropy captured in the wave-function ansatz.
Thus, we anticipate FATTN will be an efficient practical numerical tool in the future simulation of two dimensional quantum many-body systems on a finite lattice.

%In our simulations of the transverse Ising model, we see a big improvement than ATTN and TTN in the single-layer TTN and the multi-layer TTN for the transverse Ising model. We also find that the most efficient way is to put the disentangler just at the bottom layer. However, there are some problems. The entropy of our tensor network structure is limitedly scaling as $S=cLlog(4)+log(D)$, where c is a constant decided by the tensor network structure. And we may find that it is more efficiency to increase c rather than $D$ as the entropy increase linear with $c$. But we find that it is more efficiency to increase $D$ in our simulations of Fig.~\ref{mul_result}. However, because we just calculate the lattice size of $8\times 8$ and only use the transverse Ising model to benchmark our results, we can not make sure about this strange phenomenon until we have calculated other larger lattice sizes and other hamiltonian models.

%In conclusion, we want to mention that the FATTN, having a wonderful computational cost $O(D^4d^4)$ while encoding the entanglement-entropic area law as much as possible, is the most efficient method for simulations of the high dimensional systems. But we need to specify that we haven't implemented any symmetries \cite{PhysRevB.83.115125,PhysRevB.86.195114,WEICHSELBAUM20122972} in our tensor network structure which can further improve the numerical precision and efficiency of our tensor network methods. So, for the next step, we want to implement symmetries in our tensor network structure to get a better performance. 

\begin{acknowledgments}
The calculation in this work is carried out with Quimb \cite{gray2018quimb} and TensorNetwork \cite{roberts2019tensornetwork}.
This work is supported by a start-up fund from School of Physics and Astronomy in Shanghai Jiao Tong University.

\end{acknowledgments}

\bibliography{FATTN}

\begin{thebibliography}{50}
\bibitem[{foo({\natexlab{a}})}]{foot2}
\bibinfo{note}{MERA is constructed with isometries and disentanglers whose
  definition will be introduced in the main text later.}
\bibitem[{foo({\natexlab{b}})}]{foot3}
\bibinfo{note}{A detailed analysis of entanglement entropy of FATTN can be
  found in the appendix. There is a requirement on bond dimension $D$ to
  fulfill the $L\log(d^2)$ scaling. In this sense, FATTN is designed to study
  finite systems.}
\bibitem[{foo({\natexlab{c}})}]{foot1}
\bibinfo{note}{Actually TTN can be constructed from arbitrary rank-3 tensors.
  But with isometry, the construction becomes trivial.}
\end{thebibliography}

\appendix

\section{The entanglement entropy of FATTN}
\begin{figure}[t]
	\includegraphics[width=80mm]{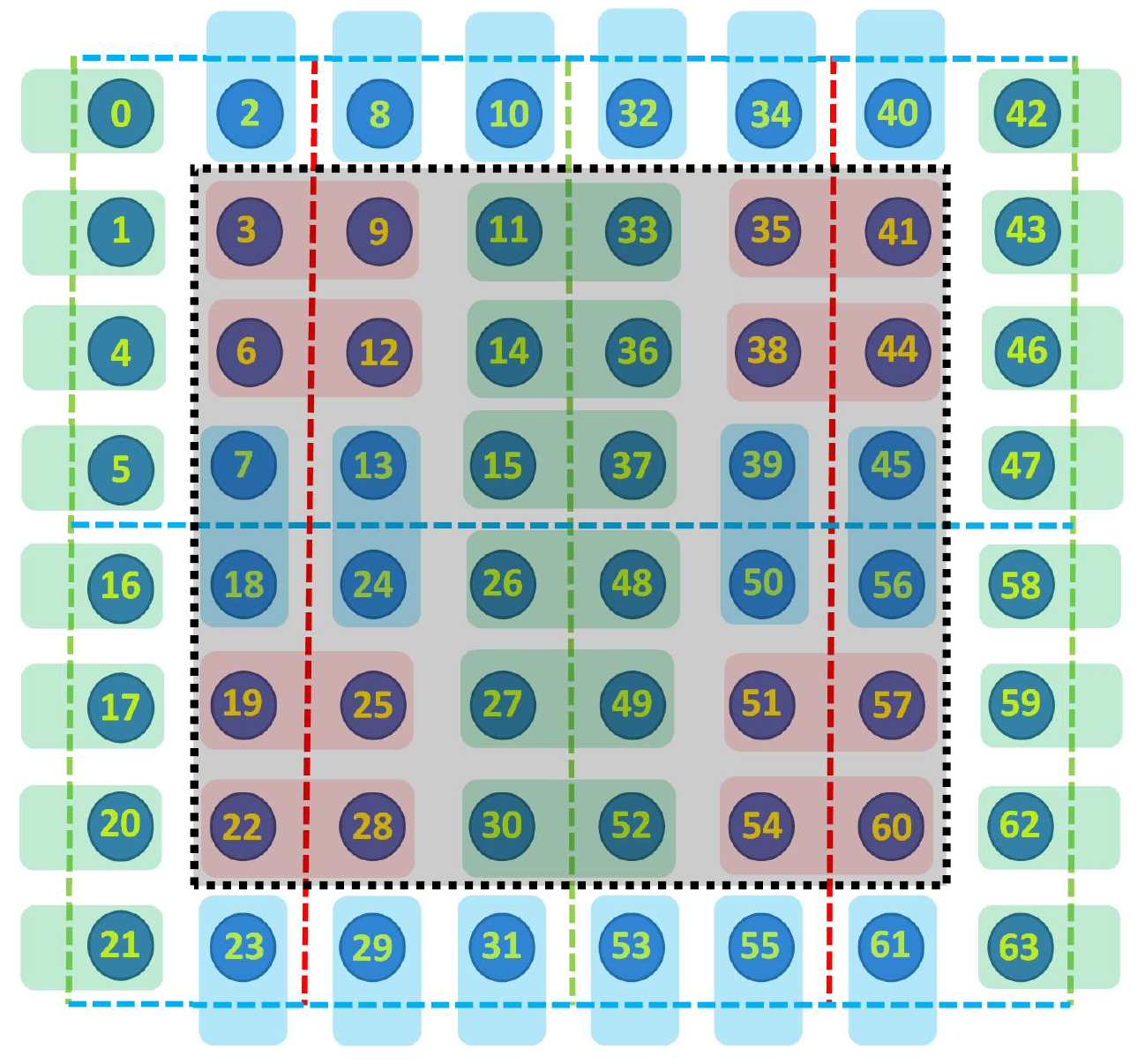}
	\includegraphics[width=80mm]{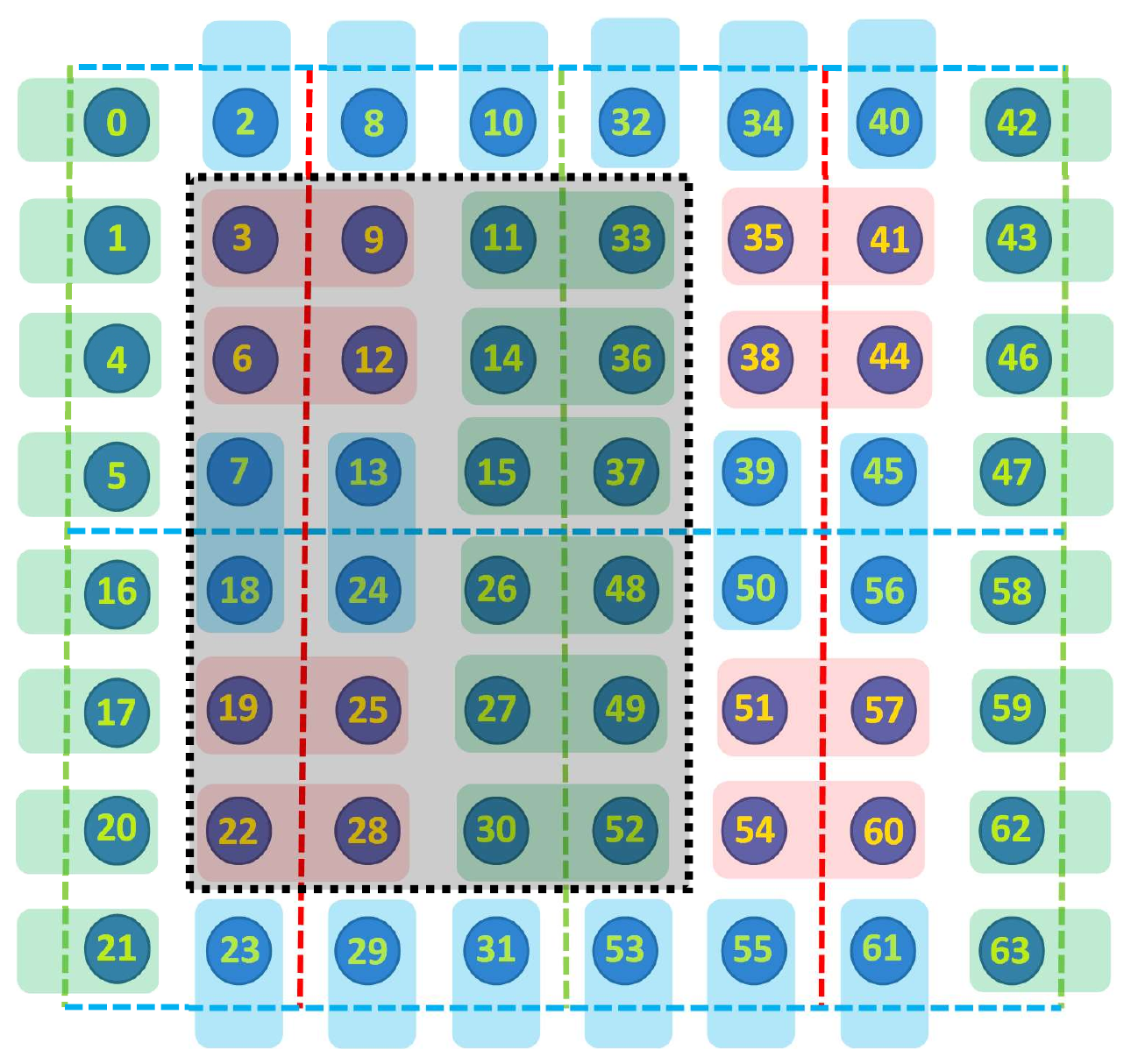}
	\caption{Two cuts for a $8 \times 8$ lattice which is not discussed in the main text. There is no disentangler across these cuts in FATTN.}
	\label{boundray2}
\end{figure}

\begin{figure}[b]
	\includegraphics[width=80mm]{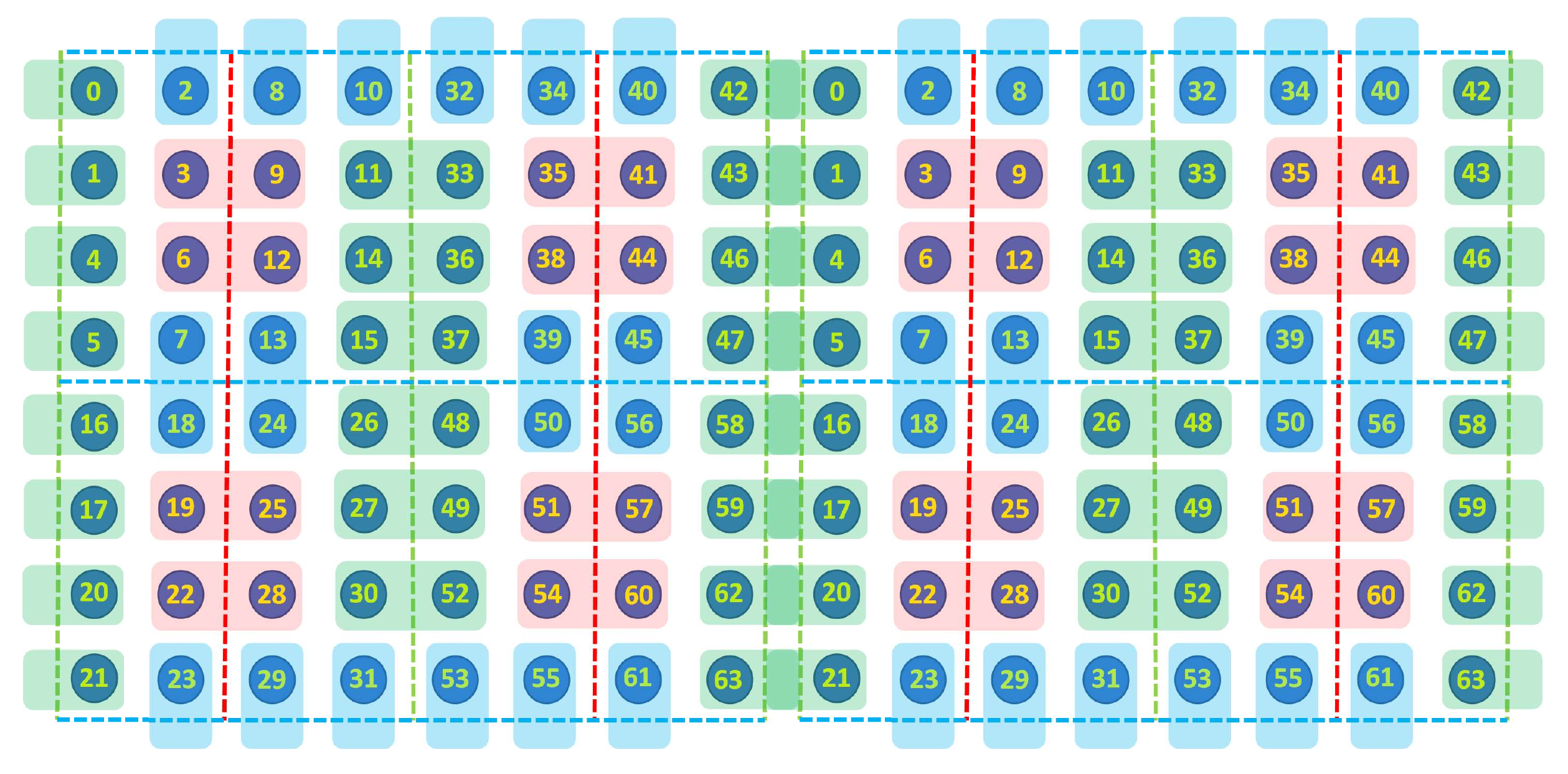}
	\caption{The way to place disentanglers of FATTN on a $8 \times 16$ lattice following the strategy in $8 \times 8$ lattice in Fig.~\ref{NewTTN_dis2}.
		This strategy can be repeated for the placement of disentanglers in FATTN for larger systems.}
	\label{FATTN16}
\end{figure}

In this section, we give a detailed analysis on the entanglement entropy the FATTN can encode. In the main text, we show that the entanglement entropy along the colored dash line in the Fig.~\ref{NewTTN_dis2}
is $L\log(d^2)$.
And the requirements for the bond-dimension to encode the entropy of $L\log(k)$ along the colored boundaries are:
\begin{equation}
	\begin{cases}
		D_g=d^4(\frac{k}{d^2})^{16}\\
		D_b=d^6(\frac{k}{d^2})^8\\
		D_r=d^4(\frac{k}{d^2})^4\\
	\end{cases}
	\label{cons_bond}
\end{equation}
Now, we will show that if these conditions are satisfied, the requirements for most of the other cuts will be also satisfied.

We firstly consider a $8\times 8$ lattice. For the black area shown in the upper panel of Fig.~\ref{boundray2} where we do not place any disentangler. The boundary length of the cut is $20$.
So in order to encode an entropy of $\log(k^{20})$ the requirement for  {bond-dimension} is:
\begin{equation}
	k^{20}=D^4d^{20}\Rightarrow D=d^5(\frac{k}{d^2})^5
	\label{K20}
\end{equation}

So once again, we need $k\leq d^2$ if we want to use a small bond-dimension $D$ to encode the entanglement-entropic area law of $L\log(d^2)$.

For the bipartition shown in the down panel of Fig.~\ref{boundray2}, the requirement for  {bond-dimension} is:
\begin{equation}
	k^{16}=D^2d^{16}\Rightarrow D=d^8(\frac{k}{d^2})^8
	\label{K16}
\end{equation}
We still need $k\leq d^2$ if we want to use a small bond-dimension $D$ to encode the entanglement-entropic area law of $L\log(d^2)$. For other cuts whose boundary is smaller or there are disentanglers across them,
the requirement is also satisfied. So, the FATTN can efficiently encode an entanglement-entropic area law of $L\log(d^2)$ for a $8\times 8$ lattice.

%For larger systems, it is quite easy to analysis.
We can iteratively use the strategies for $8 \times 8$ lattice shown in Fig.~\ref{NewTTN_dis2} to analyze larger lattices. 
In Fig.~\ref{FATTN16}, we take a $16\times 8$ lattice as an example. For a general finite system,
%considering \eqref{K20}, \eqref{K16} and the contributions of disentanglers, we can write:
the requirement for {bond-dimension} of an area law as $L\log(k)$ is 
\begin{equation}
	k^L=D^{L/c}d^Ld^{pL}\Rightarrow
	D=d^{c(1-p)}(k/d^2)^c
\end{equation} where $pL$($p  \in [0,1]$) is the number of disentanglers across the cut and $c$ is a cut dependent constant.
Because we place disentanglers across the cuts where c is large, which means $p$ is close to $1$ if $c$ is large.
So for most of the cuts $c(1-p)$ is relatively small which means for these cut the captured entanglement entropy scales $L\log(d^2)$.
We can also increase the dimension of the relevant bonds for these cuts with large $c(1-p)$ but maintain a relatively small D for other bonds.  

%And there are only a little of special cases that $c(1-p)$ is very large. But it won't influence our result dramatically in our actual simulations. The main reason is that it is the maximum entropy we can encode($S=\sum_{i=1}^x-p_i\log(p_i)\leq log(x)$). In actual simulations, the spectrum of the reduced density matrix affects the value of entropy dramatically(Otherwise, ATTN, encoding an entropy of $L\log(2)$, is enough for most of the quantum many body system). And we merely want to use disentanglers to maximally reduce the value of the bond-dimension $D$ \cite{PhysRevLett.126.170603}. 

%In conclusion, we can say that the FATTN can efficiently encode an entropy of $L\log(4)$ and must be better than the ATTN from the point of entropy.

\end{document}